
\input amssym.def
\input amssym
\font\de=cmssi12

%


\ifx\UsualIsLoaded\undefined
\let\UsualIsLoaded=\relax		

\font\fourteenrm=cmr12  scaled \magstep1
\font\fourteenbf=cmbx12 scaled \magstep1
\font\fourteentt=cmtt12 scaled \magstep1
\font\fourteensl=cmsl12 scaled \magstep1
\font\fourteensy=cmsy10 scaled \magstep2
\font\fourteeni=cmmi12  scaled \magstep1
\font\fourteenit=cmti12 scaled \magstep1
\font\fourteensc=cmcsc10 scaled \magstep2
\font\fourteenbit=cmssi12 scaled \magstep1

\font\twelverm=cmr12
\font\twelvebf=cmbx12
\font\twelvett=cmtt12
\font\twelvesl=cmsl12
\font\twelvesy=cmsy10 scaled \magstep1
\font\twelvei=cmmi12
\font\twelveit=cmti12
\font\twelvesc=cmcsc10 scaled \magstep1
\font\twelvebit=cmssi12
\font\tenrm=cmr10
\font\tenbf=cmb10 
\font\tentt=cmtt10 
\font\tensl=cmsl10 
\font\tensy=cmsy10 
\font\teni=cmmi10 
\font\tenit=cmti10 
\font\tensc=cmcsc10 
\font\tenbit=cmssi10
\font\ninei=cmmi9 
\font\ninerm=cmr9

\font\ninesy=cmsy9 
\font\eighti=cmmi8
\font\eightrm=cmr8

\font\eightsy=cmsy8
\font\seveni=cmmi7 
\font\sevenrm=cmr7
\font\sevensy=cmsy7
 
%
%

%

\def\tenpoint{
\def\rm{\fam0\tenrm}
\textfont0=\tenrm \scriptfont0=\eightrm \scriptscriptfont0=\sevenrm
\textfont1=\teni \scriptfont1=\eighti \scriptscriptfont1=\seveni
\textfont2=\tensy \scriptfont2=\eightsy \scriptscriptfont2=\sevensy
\textfont3=\tenex \scriptfont3=\tenex \scriptscriptfont3=\tenex

\textfont\itfam=\tenit
\def\it{\fam\itfam\tenit}

\textfont\slfam=\tensl
\def\sl{\fam\slfam\tensl}

\textfont\bffam=\tenbf
\def\bf{\fam\bffam\tenbf}

\textfont\ttfam=\tentt
\def\tt{\fam\ttfam\tentt}

\def\sc{\tensc}

\def\bit{\tenbit}

\normalbaselineskip=12pt

\setbox\strutbox=\hbox{\vrule height10pt depth4pt width0pt}%
\normalbaselines\rm}


\def\twelvepoint{
\def\rm{\fam0\twelverm}
\textfont0=\twelverm \scriptfont0=\ninerm \scriptscriptfont0=\sevenrm
\textfont1=\twelvei \scriptfont1=\ninei \scriptscriptfont1=\seveni
\textfont2=\twelvesy \scriptfont2=\ninesy \scriptscriptfont2=\sevensy
\textfont3=\tenex \scriptfont3=\tenex \scriptscriptfont3=\tenex

\textfont\itfam=\twelveit
\def\it{\fam\itfam\twelveit}

\textfont\slfam=\twelvesl
\def\sl{\fam\slfam\twelvesl}

\textfont\bffam=\twelvebf
\def\bf{\fam\bffam\twelvebf}

\textfont\ttfam=\twelvett
\def\tt{\fam\ttfam\twelvett}

\def\sc{\twelvesc}
\def\bit{\twelvebit}

\normalbaselineskip=14pt

\setbox\strutbox=\hbox{\vrule height10pt depth4pt width0pt}%
\normalbaselines\rm}


\def\fourteenpoint{
\def\rm{\fam0\fourteenrm}
\textfont0=\fourteenrm \scriptfont0=\twelverm \scriptscriptfont0=\tenrm
\textfont1=\fourteeni \scriptfont1=\twelvei \scriptscriptfont1=\teni
\textfont2=\fourteensy \scriptfont2=\twelvesy \scriptscriptfont2=\tensy
\textfont3=\tenex \scriptfont3=\tenex \scriptscriptfont3=\tenex

\textfont\itfam=\fourteenit
\def\it{\fam\itfam\fourteenit}

\textfont\slfam=\fourteensl
\def\sl{\fam\slfam\fourteensl}

\textfont\bffam=\fourteenbf
\def\bf{\fam\bffam\fourteenbf}

\textfont\ttfam=\fourteentt
\def\tt{\fam\ttfam\fourteentt}

\def\sc{\fourteensc}
\def\bit{\fourteenbit}

\normalbaselineskip=16pt

\setbox\strutbox=\hbox{\vrule height10pt depth4pt width0pt}%
\normalbaselines\rm}

%


%

\fi			

\def\myquad{\hskip .75em\relax}
\def\inv{^{-1}}
\def\reals{{\Bbb R}}

\def\integers{{\Bbb Z}}
\def\complexes{{\Bbb C}}
\def\Ga{\Gamma_\alpha}
\def\Gb{\Gamma_\beta}
\def\Go{\Gamma_1}
\def\Gt{\Gamma_2}
\def\Gr{\Gamma_3}

\def\Ham{Hamiltonian}

\def\ba{{\bf a}}
\def\bK{{\bf K}}
\def\bs{\bar\sigma}
\def\PP{{\cal P}}
\def\AA{{\cal A}}
\def\LL{{\cal L}}
\def\XX{{\cal X}}
\def\circle{{\bf S}^1}
\def\diffeos{diffeomorphisms}
\def\diffeo{diffeomorphism}
\def\frac#1#2{{#1\over #2}}

\def\hAz{{\hat A_0}}
\def\hH{{\hat H}}

\def\homeos{homeomorphisms}
\def\homeo{homeomorphism}
\def\ie{i.e. \kern -2 pt }

\def\mydots{\ \dots \ }
\def\pos{periodic orbits}
\def\po{periodic orbit}

\def\pp{periodic point}
\def\raw{\rightarrow} 
\def\section#1{\medskip\noindent{\bf #1:}\medskip}
\def\s{\sigma}
\def\torus{S^1\times S^1}
\def\xa{x_\alpha}
\def\ya{y_\alpha}
\def\za{z_\alpha}
\def\zb{z_\beta}
\def\Za{Z_\alpha}
\def\zaa{z_\alpha^\ast}
\def\Zb{Z_\beta}

\def\cylinder{{S^1\times \reals}}
\def\hZ{{\hat Z}}
\def\IM{{\rm Im}}
\def\RE{{\rm Re}}
\def\Ro{R^o}
 
\def\so{\sigma_1}
\def\st{\sigma_2}
\def\sr{\sigma_3}
\def\bso{{\bar\sigma_1}}
\def\bst{{\bar \sigma_2}}
\def\bsr{{\bar \sigma_3}}
\def\TN{Thurston-Nielsen}
\def\physb{${\bf b}$}
\def\bb{{\bf b}}
\def\trivb{{\bf b}_{triv}}
\def\wonko{weakly equivalent}
\def\hB{{\hat B}}
\def\fo{finite order}
\def\pA{pseudoAnosov}
\def\adhom{advection homeomorphism}
\def\BH{Bestvina-Handel}
\def\X{{\cal Y}}
\def\torus{S^1 \times S^1}
\def\TT{{\Bbb T}}

\input epsf
\fourteenpoint

\centerline{\bf Topological fluid mechanics of point vortex motions}
\twelvepoint
\bigskip
{\narrower

\noindent {\bf Philip Boyland}\footnote{$^1$}{To whom correspondence should be
addressed: {\tt boyland@math.ufl.edu}},
 Department of Mathematics,
 University of Florida,
 358 Little Hall,
 Gainesville, FL 32611-8105
\medskip
\noindent{\bf Mark Stremler} and {\bf Hassan Aref}, 
Department of Theoretical and Applied Mechanics,
University of Illinois,
104 South Wright Street,
Urbana, IL 61801
\bigskip

{\bf Abstract:}
Topological techniques are used to study the motions
of systems of point vortices in the infinite plane, in
singly-periodic arrays, and in  doubly-periodic lattices.
The reduction of each system using its symmetries is described in
detail. Restricting to three vortices with zero
net circulation, each reduced system is described by a one degree of freedom
Hamiltonian. The phase portrait of this reduced system is subdivided
into regimes using the separatrix motions,  and
a braid representing the topology of all vortex motions
in each regime is computed. This braid also describes the isotopy class of
the advection \homeo\ induced by the vortex motion. The \TN\
theory is then used to analyse these isotopy classes, and in certain
cases strong conclusions about the dynamics of the advection
can be made.

}
\bigskip

\noindent{\bf \S1 Introduction.}

The modelling of incompressible flow at high Reynolds number as a potential
flow with embedded vortices has repeatedly proven  useful both for
analytical and numerical purposes. The subject has
inspired numerous reviews, each stressing different
aspects of the field. The articles by 
Aref [Af1], Chorin [C1-3], Leonard [L], Majda
[Mj], Moffatt \& Tsinober [MfT], Pullin \& Saffman [PS],  Saffman \& Baker
[SB], Saffman [S], Sarpkaya [Srp], Shariff \& Leonard [SL], and
Zabusky [Z] provide a representative sample.  We are concerned here with the
further simplification of modelling a two-dimensional flow by a finite
collection of point vortices.  While this system is admittedly highly idealised,
it has found application and, to some extent, experimental verification
starting with von K\'arm\'an's analysis in 1912 of the instability of the
vortex street wake behind a cylinder and Onsager's 1949 explanation of the
emergence of large coherent vortices in two-dimensional flow through an
'inverse cascade' of energy. (See the literature cited for further details.)

In this paper we study the evolution of three kinds of point vortex
systems, in the infinite plane, in singly-periodic arrays, and in
doubly-periodic lattices. An array of point vortices is useful in modelling
flows in shear layers, wakes and jets.  The motivation for studying vortex
lattices comes from the problem of two-dimensional turbulence, a paradigm
of atmospheric and oceanographic flows, and to a lesser extent from the
study of vortex patterns formed in superfluid Helium.

Kirchhoff recognised that the evolution of $N$ point vortices could be
formulated within the framework of classical mechanics as an $N$-degree of
freedom Hamiltonian system. In many ways point vortices are the
fluid mechanical analog of point masses evolving under the mutual
interaction of Newtonian gravity. As with the $N$-body problem, point
vortex systems have continuous symmetries that give rise to integrals of
motion. In the case of zero net circulation (a case of considerable physical
interest), the two integrals arising from translation invariance are in
involution. Thus for three vortices, the Hamiltonian system is completely
integrable. Fixing a value of these integrals and factoring out by the
corresponding group action gives rise to a new Hamiltonian system 
with two less degrees of
freedom. For three vortex systems, this process of Jacobi-Poincar\'e
reduction yields a single degree of freedom system.
Vortex arrays and lattices also have a discrete symmetry coming from the
array or lattice structure.
For rational circulation ratios these result in a discrete symmetry of the
reduced system whose phase space is consequently a cylinder or torus.
Symmetries and reduction are explained in detail in \S 3.

Since the reduction is accomplished by  factoring out the
translational  invariance, trajectories in the reduced system 
describe the evolution of the shape and
orientation of the triangle spanned by the three vortices. Because the
reduced system has one degree of freedom 
the generic bounded orbit is periodic. This implies that the corresponding
triangle of vortices resumes its shape and orientation after one period.
However, the triangle may be translated. This translation vector is a
dynamic phase of the type that arises when the evolution of a full system
is reconstructed from a periodic trajectory of a reduced system. 
(For a general introduction to reduction and dynamic phases see 
[MZ1] and [MZ2]. Further references are given in \S 3.)

Having performed the reduction on a three-vortex system, the dynamics of
the vortices can be understood by analysing a one degree of freedom
Hamiltonian system (see [Af2], [AS1] and [SA]). 
In such systems the phase portrait is organised by the
saddle points and the connections between them, known as separatrices (see
Figures 3.1 and 3.2). For vortices the reduced Hamiltonian will also have
singularities or poles arising from collisions, or more properly, from the
superposition of pairs of vortices as collisions do not
occur when there is zero total circulation.
The separatrices naturally divide the phase space into regions (called
regimes in this paper) in which one would expect the corresponding fluid
motions to share certain dynamical characteristics. These characteristics
are topological and are precisely described using Artin's braid group. The
assignment of a braid to a vortex motion is accomplished by visualising the
vortex motion in  a three dimensional space-time with the two space
directions in the usual $x y$-plane and 
time going upwards. The motion in
the vortex frame is periodic, and so the strands representing the motions
in this frame have the same initial and final position. The translation
into a mathematical braid in Artin's group is accomplished by projecting
onto a fixed plane which is orthogonal to the $x y$-plane.
This construction is described in \S4 where it
is shown that all the vortex motions arising from the same regime are
described by the same braid. Examples  in the infinite plane
and in an array are worked out in detail.

As a Hamiltonian system the vortex evolution takes place in a
$2n$-dimensional phase space.  However, from a fluid mechanical
perspective it is much more natural  to
view the motion as $N$ trajectories in the plane (or the cylinder or torus
in the array and lattice cases). Since the vortices carry concentrated
vorticity, they generate a velocity field in the surrounding fluid. The
evolution of passive tracer particles in this velocity field is called
advection. In mathematical terms, advection is the evolution according to
solutions of the differential equation corresponding to the vector field of
velocities. Returning to the case of three vortices with zero net circulation,
we pass to the vortex frame to eliminate the dynamic phase
and so obtain a periodic motion of the vortices. In this frame the vortices
generate a periodic velocity field, and so we define the {\de advection
\homeo } as the Poincar\'e map obtained by advecting for one period in this
frame. Thus iterates of the advection \homeo\ describe the dynamics of
advection in the vortex frame (see \S6.1).

In \S5.1 the braid of a regime is connected to the corresponding
advection \homeos\ using the notion of isotopy. Two \homeos\ are isotopic
if one can be obtained from the other by a continuous deformation. 
In addition to its role
as a description of the topology of a periodic motion of $N$ points in the
plane, a braid on $N$ strands also naturally describes an isotopy class of
homeomorphisms on the $N$-punctured plane.
Since all the motions in a regime have the same braid, they all generate
isotopic advection \homeos. This knowledge is then applied using 
\TN\ theory of surface \homeos.

The \TN\ theory (described in \S 5.2 and \S5.3) contains a classification
theorem for isotopy classes of surface \homeos. In the case of a \pA\ class
the theory gives dynamical behaviour that must be present in every
\homeo\ in the isotopy class. This information includes a lower bound for
such quantities and structures as the growth rate of the number of \po\ points
as the period grows, the topological entropy, the growth rate of the length
of topologically nontrivial material lines, and the topology of invariant
manifold templates. In particular, any \homeo\ in a pA class will be
chaotic under any of the usual definitions of the word. 
For three vortices in the plane
with zero net circulation, the advection \homeos\ are never in \pA\ classes,
but for vortex arrays and lattices, \pA\ classes are common. Examples
are discussed in detail in \S6.2 and \S6.3.

Because we believe the methods described in this paper have general
applicability, an effort has been made to make the paper accessible to a
general scientific reader, perhaps lacking expertise in topology and/or fluid
mechanics. A fair amount of expository material has been included. In
certain cases (eg. \S4.1 and \S5.1) basic material needed in a form not
available in the literature is described in some mathematical detail. The
reader may prefer to skip these sections in a first reading.

\medskip

\noindent{\bf \S 2 Equations of motion and \Ham s.}

We shall be considering the evolution of three
types of systems of vortices:  finite collections in the infinite plane,
singly-periodic arrays and doubly-periodic lattices.
The names ``array'' and ``lattice'' will be reserved
for the singly- and doubly-periodic systems, respectively.
Both real and complex notation will be used for points in
the plane with $z = x + i y$ in all cases.
A singly-periodic array  of $N$ vortices of period (or
width)  $L$ is a collection of vortices in the plane
with  exactly $N$ vortices in each 
vertical strip of width $L$ and 
if there is a vortex at position  given by a complex number 
$\za$ there is also one at $\za + nL$ for all 
integers $n$. The definition of a vortex lattice 
requires a pair of linearly independent
complex numbers $\omega_1$ and $\omega_2$ (the {\de 
half-periods}) 
that determine the double-periodicity. 
A doubly-periodic lattice of $N$ vortices 
with lattice structure generated by $\omega_1$ and $\omega_2$
is a collection  of point vortices that 
has exactly $N$ vortices in each fundamental
parallelogram determined by $2 \omega_1$ and $2 \omega_2$, 
and if there is a vortex at position $\za$ there is also one 
at $\za + n_1 2 \omega_1 + n_2 2 \omega_2$ for all 
 integers $n_1$ and $n_2$.
Eventually we shall describe  vortex arrays
or lattices  by  singly- or doubly-periodic coordinates, 
or equivalently, as $N$ vortices on
a cylinder or torus.  However, initially 
the arrays or lattices are specified by a collection of 
$N$ distinguished points.  Thus, for example,
$(z_1 + L, z_2, z_3)$ represents a different array than
$(z_1, z_2, z_3)$. Note also that our definition of 
lattice requires that it maintain the same periodicity for
all time, so a uniformly rotating system
is not a lattice in our sense.
\medskip

\noindent{\bf \S 2.1 Equations of motion for the vortex systems.}
 The $\alpha^{th}$  point vortex has  a constant circulation given by
the real number $\Gamma_\alpha$ and  
a position at time $t$ given by the complex number
$\za(t)$. Using an asterisk 
denotes the complex conjugate,  the $N$ differential
equations that describe the
evolution of the vortex systems are
$${d{\zaa}\over dt} = 
{1\over 2 \pi i} \sum_{\beta\not = \alpha}
\Gamma_\beta \;\phi_\XX(\za - \zb) \eqno{(2.1})$$
for $\alpha=1, \mydots, N$.
Using $\XX=\PP$ for the infinite plane, $\XX=\AA$ for  arrays, 
and $\XX=\LL$ for lattices, the complex-valued function $\phi_\XX$ in
the various cases  is
$$\eqalign{
\phi_\PP(z) &= {1\over z}\cr
\phi_\AA(z) &= {\pi \over L} \cot({\pi\over L} z)\cr
\phi_\LL(z) &= \zeta(z) + \delta z  - {\pi\over \Delta}
z^\ast \cr}\eqno{(2.2}) $$
where in the lattice case,
 $\zeta(z) = \zeta(z; \omega_1, \omega_2)$ is the
Weierstrass  zeta function,  
$\Delta$ is the area of a fundamental parallegram, and
$\delta = {\pi \over \Delta} - 
{\zeta(\omega_1)\over \omega_1}$. Note that in each case the
function $\phi$ is odd and has the required periodicity;
$\phi_\LL(z + n L) = \phi_\LL(z)$ and 
$\phi_\AA(z + 2 n_1 \omega_1 + 2 n_2 \omega_2) = \phi_\AA(z)$
(the latter identity uses the Legendre relation for the
Weierstrass zeta function).

In the presence of the evolving vortex system, a
passive particle with position given by
$z(t)$  advects according to
$${d{z^\ast}\over dt} = {1\over 2 \pi i} 
\sum_{\beta = 1}^n \Gb\;\phi_\XX(z -  \zb(t))\eqno{(2.3})$$
with the function $\phi_\XX$ the same as above.
One may think of the passive particle as a vortex
with zero circulation, and so the advection
problem is the analog of the Newtonian $(N+1)$-body problem.

The equations of motion for the plane and array
case are well known.   For a derivation of the equations
for vortex lattices see [O]  and [SA].
The function $\phi$ is usually
thought of as representing the contribution of a single vortex.
It has a simple pole at the position of the vortex and thus
contributes a delta function to the curl of the velocity
field. Note, however,  that in the lattice case the
function $\phi_\AA$ is not  meromorphic.
This is a reflection of a property of
doubly-periodic meromorphic (\ie elliptic) 
functions, namely,  the residues of the
poles in a fundamental parallelogram must sum to zero
(equivalently, by Greens' Theorem, a doubly-periodic
vector field must have zero net curl in each fundamental
parallelogram).
Thus the concept of a single doubly-periodic vortex or of a 
lattice (as we have defined it) with non-zero net circulation requires
additional consideration. On  the other hand, if
$\sum \Ga = 0$, as will be assumed below, the right hand side
of (2.3) is an elliptic function of $z$ and one
may speak without contradiction 
of the effect of the entire lattice on a passive particle.

\medskip
\noindent{\bf \S 2.2 The Hamiltonian framework.}
It is well known that the equations of motion for
systems of point vortices can be put into the 
framework of classical Hamiltonian mechanics.
For an $N$-vortex system
the phase space is $\complexes^N - \Upsilon_\XX$ where
$\Upsilon_\XX$ is the {\de collision set} defined in the various
cases as
$$\eqalign{
\Upsilon_\PP(z) &= \{ (z_1, \mydots, z_N) \in \complexes^N :
z_i = z_j \ \hbox{for some} \ i \not = j \} \cr
\Upsilon_\AA(z) &= \{ (z_1, \mydots, z_N) \in \complexes^N :
z_i = z_j + nL \ \hbox{for some} \ i \not = j, n\in\integers \}\cr
\Upsilon_\LL(z) &= \{ (z_1, \mydots, z_N) \in \complexes^N :
z_i = z_j + 2n\omega_1 + 2 m \omega_2 \ \hbox{for some} \ i \not = j, 
  n,m\in\integers\}.\cr} $$
The $x$ and  $y$ positions of each vortex, $\xa$ and $\ya$,
are conjugate variables, so the dynamical system
is $2 N$-dimensional. The Hamiltonian system has
$N$ degrees of freedom, but there is no  configuration space
in the usual sense. Put in geometric language,
the Hamiltonian is defined on a $2 N$-dimensional symplectic
manifold that is not a cotangent bundle. 

In all cases the real-valued Hamiltonian takes the form
$$
H_\XX(z_1, \mydots, z_N) = -{1\over 4 \pi } 
\sum {}^\prime \Ga\;\Gb\; \Phi_\XX(\za - \zb)\eqno{(2.4)}
$$ 
where the sum is over all $\alpha$ and $\beta$, and the
primed summation symbol indicates that the case $\alpha = 
\beta$ is excluded. The function 
$\Phi_\XX$ (essentially the real part of the
antiderivative of $\phi_\XX$)
in the various cases is
$$
\eqalign{\Phi_\PP(z) &= \log(|z|)\cr
\Phi_\AA(z) &=  \log(|\sin({\pi \over L}z)|)\cr
\Phi_\LL(z) &= \log(|\sigma(z)|) + \hbox{Re}({\delta z^2\over 2})
- {\pi \over 2 \Delta} z z^\ast\cr} \eqno{(2.5)}
$$
with $\sigma(z)$ the Weierstrass sigma function with half periods 
$\omega_1$ and $\omega_2$.
 The equations of motion (2.1) then have the form 
$$\Ga {d \zaa \over dt} = {\partial H \over \partial \ya}
+ i {\partial H \over \partial \xa}.
\eqno{(2.6)} $$
The presence of the factor $\Ga$ on the left hand side indicates 
that in order to put the equations in the proper Hamiltonian form
we need to change coordinates or else use a slightly 
nonstandard symplectic form.
We choose the latter course and adopt the 
form 
$$\sum \Ga\; dx_\alpha\wedge d y_\alpha,\eqno{(2.7)}$$
and so  the skew-symmetric matrix representing the form is 
$$\Omega = \pmatrix{0 & -\Gamma\cr
                   \Gamma & 0}$$
with $\Gamma = \hbox{diag}(\Ga)$.  Thus if 
$W = (x_1, \mydots, x_N, y_1, \mydots, y_N,)$, 
the equations of motion (2.2) become
$$\Omega\;{d W\over dt} =  \nabla H$$ 
with $\nabla$  the real gradient.

It is clear from equations (2.1) that  
the complex quantity $J = \sum \Ga\za$ is a constant
of the motion. This integral is associated with the
invariance of the Hamiltonian under simultaneous planar
translation of all the vortices. 
Writing $J = Q + i P$ with $P$ and $Q$ real-valued,
and using the symplectic form (2.7), the Poisson bracket 
of $P$ and $Q$ is
 $ \{ P, Q\} = (\nabla P)^T\; (\Omega\inv)^T\; \nabla Q 
= \sum \Ga$. 
Throughout most of this paper we will be assuming that
$\sum \Ga = 0$, and so $P$ and $Q$ are in involution.  
The  gradients of $P$ and $Q$ are obviously independent.
Thus  in the case $N=3$ the system (2.6) 
is completely integrable. The structure of the level sets
is discussed in \S 3.5 below.

The advection equation (2.3) may also be put in 
Hamiltonian form. Assuming that a  motion $\{ \zb(t) \}$
of the $N$ vortices is given, the 
 time-dependent real-valued Hamiltonian is 
$$G_\XX(z) =  -{1\over 2\pi} 
\sum_{\beta=1}^N \Gb \Phi_\XX(z - \zb(t)).\eqno{(2.8)}$$
In   this case we use the  standard symplectic form on the plane. 
The function $G_\XX$  is more commonly called the 
{\de stream function} of the advection problem. The right
hand side of (2.8) is the velocity field of
an unsteady two-dimensional fluid motion generated by the 
evolving collection of vortices.

One can also combine the systems (2.4) and (2.8) into a single
$(N+1)$-degree of freedom system, but we do not pursue this
point of view here. 

\medskip

\noindent{\bf \S 3 Symmetry and reduction.}

The evolution of the vortex systems is certainly independent
of the choice of the origin.  This is reflected in the invariance 
 of the \Ham\ (2.4)  under simultaneous translation of 
all the vortices. 
There is a well-known process going back to
Jacobi and Poincar\'e which uses such symmetries  to reduce the 
effective dimension of a \Ham\ system. This process of
{\de reduction} proceeds by first fixing a value of the integrals
arising from the symmetries and then identifying elements in the 
level set that correspond
under the   restricted symmetry. The resulting reduced system
is \Ham\ and has one less degree of freedom for every independent
continuous symmetry. The mathematical theory of
reduction has been much developed in recent years (see, for
example, [M], Appendix 5 in [A2], or Chapter V.D in [MH]).
 We shall need only a small piece of the theory here.
Reduction for vortices in the infinite plane 
is described in greater generality in  [AR].

In addition to the continuous symmetries, vortex arrays and
lattices have  discrete symmetries that describe their periodicities. 
The utilisation of the discrete symmetries is  rather different 
than continuous ones, resulting in a change of
the topology of the phase space rather than a reduction
of dimension.
\medskip

\noindent{\bf \S 3.1 Group actions.} 
Symmetries are usually described by 
invariance under a group action. If 
$G$ is a group  (which we write additively), 
an action of $G$ on a space $X$  is a (at least continuous) 
map $G \times X \raw X$, where $g \cdot x$ is written
for the image of $(g, x)$. The action is required to satisfy 
$0 \cdot x = x$ and
$(g_1 + g_2)\cdot x = g_1\cdot(g_2 \cdot x)$. The {\de orbit} or 
{\de trajectory} through
$x$ is the image of $x$ under all the group elements,
sometimes written $G \cdot x = \{ g\cdot x : g\in G\}$.

 Given a Hamiltonian 
$H:X \raw \reals$ and an action of the  group $G$, the Hamiltonian
is said to be invariant under the group action if
$H(g\cdot x) = H(x)$ for all $x\in X$ and $g\in G$. Because of this
invariance the Hamiltonian will be constant on the orbits of the action.
Thus if we create a new space by identifying all the
points on the same orbit, we will get a new Hamiltonian function
on the quotient space. This process of collapsing orbits to points
is called {\de factoring or mod-ing} out by the group.

A group action is also generated by the collection 
of solutions to an autonomous ordinary differential
equation, $\dot x = F(x)$ with $x\in X$ (if
both $F$ and $X$ are sufficiently well behaved). 
 This  action of $\reals$ on $X$ is usually 
written  $\psi: X \times\reals \raw X$ and it satisfies 
$\partial\psi(x, t)/\partial t = F(\psi(x, t))$. In Dynamical
Systems this action is called a {\de flow}, and here
we use flow exclusively in this sense.  The evolution of 
a fluid is called a {\de fluid motion}. 
Note that if a fluid motion is steady in a region $X$, then the evolution
of passive scalars (advection) is described by a mathematical flow. 
However, if the velocity field 
$F$ depends on $t$, the motion of the fluid is not
a  flow in the sense used here.
 One can obtain a flow on a new
space  $X\times \reals$ by defining a new variable $\tau\in \reals$ 
with  $\dot \tau = 1$  and letting $\dot x = F(x, \tau)$.

For Hamiltonian systems, invariance under a group action is
closely related to integrals of motion:
if $H$ is invariant under an action of $\reals$ and this action
is the same as the flow induced by another Hamiltonian $\hat H$,
then $\hat H$ is an integral invariant of the original Hamiltonian
system. If there are $k$ independent symmetries that commute
(usually expressed
as an action of $\reals^k$), then there will be $k$ independent
integrals of motion.

\medskip
\noindent{\bf \S 3.2 Continuous and discrete symmetries of the vortex systems.}
 The \Ham\ (2.4) is invariant under simultaneous translation of 
all the vortices,  
$$
H_\XX(z_1 + \tau, \mydots, z_N+\tau)
= H(z_1, \mydots,  z_N) \eqno{(3.1)}
$$
 for $\tau \in \complexes$. 
This action of $\complexes$ (or $\reals^2$) is called 
the {\de continuous symmetry}. 
It gives rise to the integral of motion  $J := Q + i P$
defined in \S 2.2.

In addition to the continuous symmetries, vortex arrays and lattices
have a symmetry associated with their 
periodicity. For arrays,
$$H_\AA(z_1 + n_1 L, 
\mydots, z_N+ n_N L)
= H_\AA(z_1, \mydots,  z_N)\eqno{(3.2)}$$
for all collections of integers $n_1, n_2, \mydots, n_N$.  Thus the
\Ham\ is invariant under an action of $\integers^N$. In the
lattice case the \Ham\ is invariant under a 
$(\integers^2)^N$ action,  
$$H_\LL(z_1 + n_1 2 \omega_1 +  m_1 2 \omega_2, 
\mydots, z_N+ n_N 2 \omega_1 + m_N 2 \omega_2)
= H_\LL(z_1, \mydots,  z_N).\eqno{(3.3)}$$
 We  shall call these the {\de discrete symmetries}.

\medskip

\noindent{\bf \S 3.3  Reduction and reconstruction using formulae.}
From this point onward
we restrict to the case of three vortices, $N=3$.
In addition, the
sum of the circulations is assumed to be zero,  $\sum \Gb = 0$. 

If  $Z  = z_1 - z_2$, then using the definition
of $J$ and the fact  that $\sum \Ga = 0$, 
$$
\eqalign{
z_2 - z_3 &= {-J + \Go Z\over \Gr}\cr
z_1 - z_3 &= {-J - \Gt Z\over \Gr}.\cr}\eqno{(3.4)}
$$
Substituting this into the result of 
subtracting the first two equations of (2.1) yields
$$
{d  Z^\ast\over dt} = 
{-\Gr \over 2 \pi i}(\phi_\XX(Z) +
\phi_\XX({-J + \Go Z\over \Gr}) +
 \phi_\XX({J + \Gt Z\over \Gr})). \eqno{(3.5)} 
$$
This is a one degree of freedom  system with
\Ham\
$$
K_\XX(Z) = 
{\Gr \over 2 \pi }(\Phi_\XX(Z) +
{\Gr\over \Go} \Phi_\XX({-J + \Go Z\over \Gr}) +
{\Gr\over \Gt} \Phi_\XX({J + \Gt Z\over \Gr}))\eqno{(3.6)} 
$$
using the standard symplectic form (phase portraits
in a planar and array example are shown in Figures
3.1 and 3.2, respectively). Thus the two
independent integrals $Q$ and $P$ have reduced the
system by two degrees of freedom, from three to one.
However, the way in which the continuous 
symmetries have brought about the
reduction is not clear. One would expect that 
 $K_\XX$ could be obtained by substituting
(3.4) into the \Ham\ (2.4), but there is an unexplained
factor of $-{\Go\Gt\over \Gr}$. 

\midinsert
\def\epsfsize#1#2{.4\hsize}
\centerline{\epsfbox{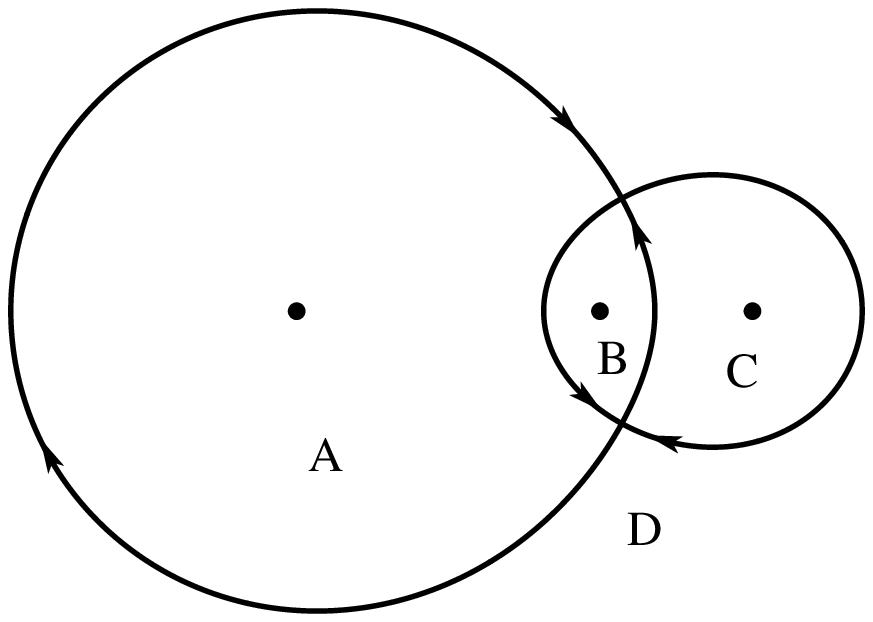}}
\medskip
{\leftskip=60pt\rightskip=60pt\parindent=0pt

{\bf Figure 3.1:}
Phase portrait of the reduced Hamiltonian system
for three vortices in the infinite plane with
circulations $1, 1/2$ and $ -3/2$. The upper case letters
label regimes discussed in \S4.3.

}
\endinsert
\midinsert
\def\epsfsize#1#2{.6\hsize}
\centerline{\epsfbox{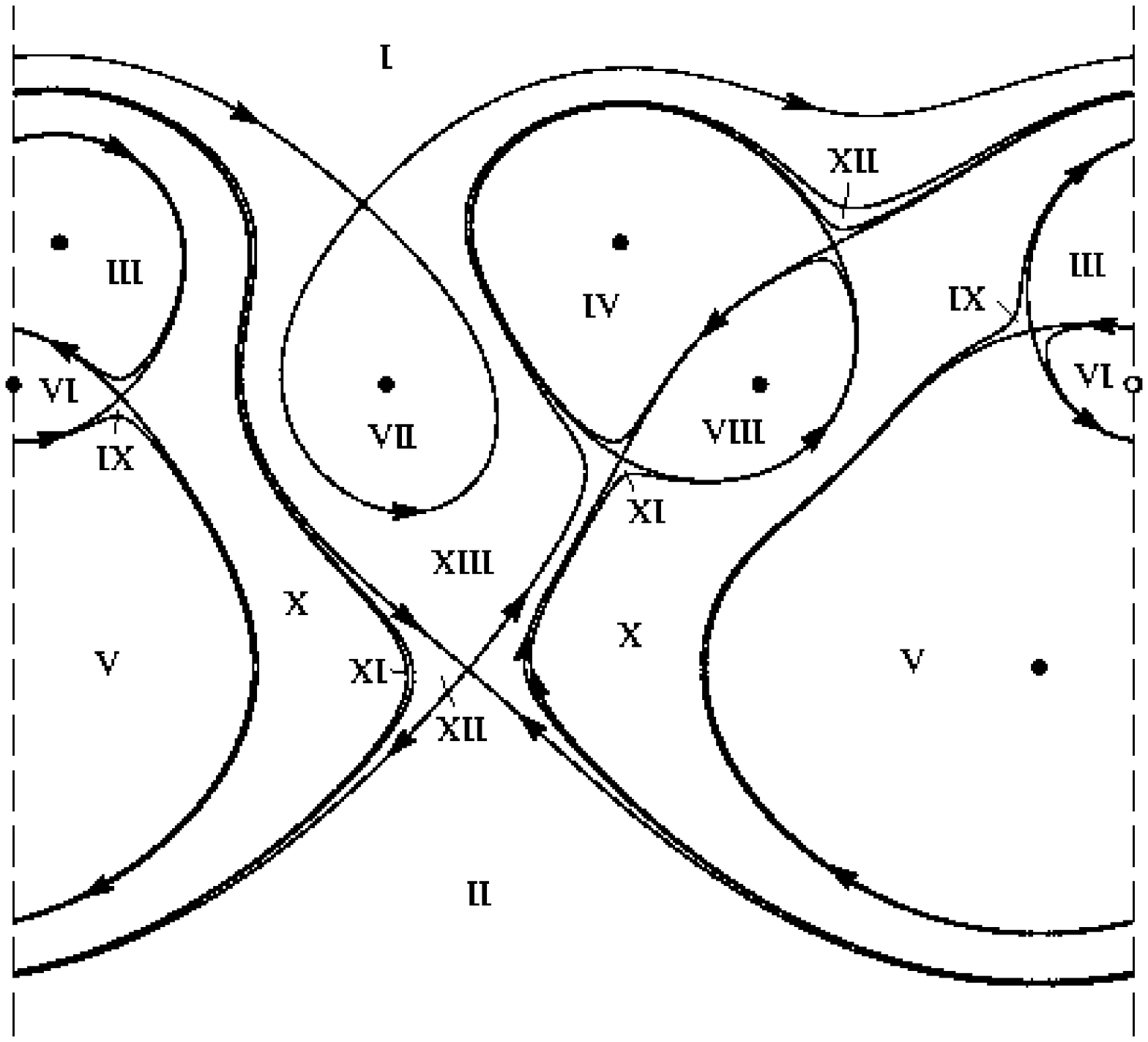}}
\medskip
{\leftskip=60pt\rightskip=60pt\parindent=0pt

{\bf Figure 3.2:}
Phase portrait of the reduced Hamiltonian system
for three vortices in a singly-periodic array with
circulations $1, 1/2$ and $ -3/2$. The Roman numerals
label regimes discussed in \S4.5.

}
\endinsert

When $\Gr$ is an integer
the discrete symmetries in  the array case are reflected in the 
invariance of $K_\AA$ under translations by $\Gr L$. Thus
 we may think of the reduced system as 
living on a cylinder of width $\Gr L$. 
But although $z_1$ and $z_2$ have a width $L$ symmetry,
why should their difference  have a 
width $\Gr L$ symmetry? These questions are answered
using the   geometric framework of the next 
subsection.

Reconstruction refers to the process of obtaining solutions
to the full equations (2.1) from the reduced one (3.5).
After fixing a value of $J$,  this may be done by ``quadrature''.
Explicitly, using  (3.4) in the differential equation (2.1) for
$z_1$,
$${d{z_1}\over dt} = 
{1\over 2 \pi i} 
(\Gamma_2 \;\phi_\XX(Z(t)) + \Gamma_3 \;\phi_\XX({-J - \Gt Z\over \Gr})).
 \eqno{(3.7})$$
Thus once an initial position $z_1(0)$ is chosen, $z_1(t)$ may
be obtained by integrating (3.7), and then $z_2(t) = z_1(t)-Z(t)$ 
and (3.4) yields $z_3(t)$.

Note that if  $Z(t)$ is given and the initial position
of the first vortex $z_1(0)$ is changed by a translation to
$\hat z_1(0) = z_1(0) + \tau$ for $\tau\in\complexes$, then the entire
motion translates by $\tau$, \ie the new solution are $\hat z_\alpha(t) 
= z_\alpha(t) + \tau$, for $\alpha = 1, 2, 3$. Also note that
if $Z(t)$ is periodic with period $P$, (3.7) shows that the vortex positions
$z_\alpha(t)$ will, in general,  not be periodic. Rather there will be
a constant $b\in\complexes$ with $z_\alpha(t+ P) = z_\alpha(t) + b$, 
for all $t$ and $\alpha$. 
The number $b$ is an example of a dynamic phase 
which is also best understood in the geometric framework of the next
subsection.

\medskip
\noindent{\bf \S 3.4 Reduction in the geometric framework.}
For most of the remainder of 
this section we  restrict  to the case of 
singly-periodic arrays and  drop the subscripts in
various notations, so $\phi = \phi_\AA$, etc. For expositional
simplicity,
 let $L=1$.

A natural starting point is the observation that
the system does not change if we
pick an integer $n$ and change  $\za$ to $\za + n $.
This means that we may treat each
$\za$ as a singly-periodic variable, \ie an element
of a cylinder $\cylinder$. This is equivalent to factoring out by the
$(\integers)^3$ symmetry and  obtaining 
a \Ham\ system with  phase space    $(\cylinder)^3 - \Upsilon$. 
Unfortunately  the integral $Q = \sum\Ga\xa$ is not well defined
as a function of the singly-periodic $x$-coordinates. 
Put another way, the 
continuous symmetry of the $\reals^2$-action is still present
in the factored system,
but it is no longer \Ham\ in character;  the 
one-parameter subgroup that translates in the
$y$-direction is not a \Ham\ flow {\it on the product of cylinders}. 
Thus we proceed by first reducing the
system by the continuous symmetries and then using
the discrete symmetries.

Fix a value of the integrals, say $J_0 = 
Q_0 + i P_0$, and let $A_0$ be the integral manifold (level set) 
$$
A_0 = \{(z_1, z_2, z_3) : \sum\Ga\;\za = J_0\}.
$$ 
It is convenient to use coordinates $(Z, Y)$ 
for $A_0$ with $Z = a_1 - a_2$ and $Y = a_1$ for
$\ba = (a_1, a_2, a_3)\in A_0$.
Since  $\sum\Ga(a_\alpha + \tau) = \sum\Ga\; a_\alpha $,
the set $A_0$  is invariant under the action of $\reals^2$ as
well as under the \Ham\ flow.  In coordinates, 
 $\tau \in \reals^2$ acts as $(Z, Y + \tau)$.
We now mod out by the $\reals^2$ action restricted to $A_0$,
 \ie we consider
$\ba$ and $\ba + \tau (1, 1, 1)$ to be the same point; 
in coordinates this just amounts to identifying 
$Y$ with $Y + \tau$ for all $\tau \in \complexes$. 
The reduced space,  denoted $R$, is thus a copy of the 
complex plane coordinatized by $Z$. 
The Meyer-Marsden-Weinstein Theorem (see [M]) assures us that
$R$ has a unique symplectic form that pulls back under the
projection $A_0\raw R$ to the
restriction of of the form $\Omega$ on $A_0$. To compute this form
on $R$, note that restricted to $A_0$, $\sum\Ga dx_\alpha =
0 = \sum\Ga dy_\alpha$.  Then using $\sum \Ga = 0$, 
$$\sum\Ga dx_\alpha \wedge dy_\alpha =
{-\Go \Gt \over \Gr} (dx_1 - dx_2)\wedge (dy_1 - dy_2) =
{-\Go \Gt \over \Gr} dZ^1 \wedge dZ^2$$
where $Z = Z^1 + i Z^2$. Thus the proper symplectic form on
$R$ is $-{\Go\Gt\over \Gr}$ times the standard one. 
The flow restricted to $A_0$ projects to a  \Ham\ flow on $R$,
and the \Ham\  is obtained by substituting
(3.4) into  the \Ham\ (2.4) of the full system. As noted
above, this substitution yields (3.6) without a factor
 $-{\Go\Gt\over \Gr}$, a factor now explained 
by the proper symplectic structure on $R$. 

The next step is to connect the \Ham\ flow on $R$ with
the full flow on $A_0$ (this is 
the  reconstruction step). 
The flow on $A_0$ may be expressed as
$$
(Z(t; Z_0), Y(t; Y_0, Z_0))\eqno{(3.8)}
$$
with the symbols after the semicolon representing the initial
conditions. Note that the evolution
of the $Z$ coordinate just depends on the $Z$ initial condition
and not on $Y$. The invariance of the \Ham\ under the
$\reals^2$-action is expressed by
$$
(Z(t; Z_0), Y(t; Y_0 + \tau, Z_0))=
(Z(t; Z_0), Y(t; Y_0, Z_0)) + \tau. \eqno{(3.9)}
$$

In more concrete terms, the variable $Z = z_1 - z_2$
in conjunction with the fixed value of $J$ 
determines the shape and orientation of the
triangle determined by the vortices. The variable  
$Y = z_1$ gives its position. Translating
this position by $\tau$ has only the effect of translating
the entire evolution by $\tau$.

Since the \Ham\ system on $R$ has just one degree of freedom, the
generic bounded orbit is periodic. Thus after some period
$P$, the shape and orientation of the triangle of vortices
will be reestablished, but in general, the entire
configuration could have translated by some amount. This
amount is the {\de dynamic phase} and is equal to
$$
b(Z_0) := Y(P; Y_0, Z_0) - Y(0; Y_0, Z_0)\eqno{(3.10)} 
$$ with $Z_0$ an element
of the chosen \po. By virtue of (3.9) this expression
is independent of $Y_0$. In geometric language this phase
represents the return map (holonomy) 
in the fiber of the bundle $A_0\raw R$ as one goes 
around the \po\ loop in $R$. The numerical value of
the phase may be obtained from $Z(t)$ by integrating
(3.7).

Thus far we have focused on the role of the continuous symmetries 
in  reducing the number of degrees of freedom of
the vortex systems. The
use of the discrete symmetries in the  array and lattice systems will be called
{\de factoring} to distinguish it from reduction. For expositional simplicity
we now assume that $\Gamma_1 = 1$ and will freely use this numerical value
as well as the symbol in various formulas. This results in no loss of
generality because
equations (2.1) show that rescaling all circulations by the 
same amount only changes the
speed (and perhaps the direction) of the  vortices, but
does not change their trajectories. 

 The main observation required in using the discrete symmetries
in the  reduction process is that
 additional symmetries can only be used if
they   preserve the level sets $A_0$,
or equivalently, the value of the integral $J$.
If $(a_1, a_2, a_3)$ is in $A_0$ there is
no guarantee that $(a_1+ n_1 , a_2 + n_2 , a_3 + n_3 )$ 
is in $A_0$ for arbitrary $(n_1, n_2, n_3)$. 
However, it is easy to find conditions  that insure this,  
namely, $\sum \Ga n_i   = 0$. The set of
all such $(n_1, n_2, n_3)$ is  a subgroup of
$\integers^3$ which we
denote $\bK$. This subgroup always contains all multiples
of $(1, 1, 1)$ and is larger exactly when 
$\Gr$ is  rational. Since translation by integer multiples
of $(1, 1, 1)$ is contained in the continuous action,
the discrete symmetries only come into play
when $\Gr$ is rational.

An element 
$ (k_1, k_2, k_3) \in \bK$ acts on $A_0$ as
$(a_1 +k_1 , a_2 + k_2 , a_3+ k_3 )$, and so it acts 
in coordinates as 
$$
(Z + (k_1 - k_2) , Y + k_1 ).\eqno{(3.10)}
$$
But note that $(k_1 - k_2) = \Gr (k_2 - k_3)$, and thus the
action of $\bK$ on $R$ just adds $\Gr$ times an integer 
to the coordinate $Z$. When $\Gr$ is irrational,
$k_1 = k_2 = k_3$, and  $\bK$ does not act on the $Z$
component. 
When $\Gamma_3 = p/q$ is  rational (in lowest terms), 
  $k_1 - k_2$ is actually 
$p$ times an integer because the $k_i$ are integers.
 Thus the induced symmetry on
$R$ is translation by $p $. Algebraically speaking, 
the map $(k_1, k_2, k_3)
\mapsto ( k_1 - k_2, k_1)$ is an isomorphism from
$\bK$ to $p\integers \oplus  \integers$. In this new form
$\bK$ acts on $R$  with just its first factor.
Thus the action of $\bK$ on $R$ induces
a  mod $p$ symmetry on the plane $R$.  
The \Ham\ on $R$ thus has a mod $p$ symmetry and so
the final factored system has  a cylinder of width
$p$ as its phase space.
Figure 3.2  shows the orbits of the reduced system
for a vortex array with $\Go = 1, \Gt = 1/2$ and $\Gr = -3/2$.
The left and right edges of the box can evidentially
be identified to get a width $3$ cylinder.

In the case of lattices, there is a 
discrete action on both the real and imaginary parts of $\za$.
Thus if $\Gr = p/q$, the factored system may be considered
a  torus with width and length $p$, \ie 
a \Ham\ on $\reals^2/p\integers^2$.

\medskip
\noindent{\bf \S3.5 Integral manifolds and dynamics.}
As noted in \S 2.2, in  the case $N=3$  the vortex systems
are completely integrable as \Ham\ systems. Orbits
of  the phase flow are constrained to simultaneous level
sets of $H, Q$ and $P$. If the phase space was compact, these
integral manifolds would be generically tori. The systems here 
do not have compact phase spaces, but the
reduction and subsequent reconstruction make the structure
of the integral manifolds clear.

In the reduced space $R$
the generic, bounded level set of $K_\XX$  is a
topological circle (dynamically a \po). If we fix one such
circle $C\subset R$, then an integral manifold $\Lambda$ of the full system
corresponding to fixing values of $H$, $Q$ and $P$ is
the preimage of $C$ under the projection $A_0 \raw R$.
Thus the generic $\Lambda$ is 
homeomorphic to $C \times \reals^2$ with the
$\reals^2$ factor coordinatized by $Y$. The flow on $\Lambda$
 factors as in (3.8). The simplest way to understand
this flow is to examine the return map of
orbits to  the  $\reals^2$ factor as one travels around 
$C$. By (3.9) and (3.10) this return map 
is translation by $b(Z_0)$ for any $Z_0 \in C$. 

In the array case when $\Gr$ is rational 
we have used the action  of the discrete 
symmetry in the first coordinate of (3.10) to factor
$R$. One also sees from (3.10) that 
the  action of the discrete symmetry
on the $Y$-component is addition of an integer.
This action can be used to factor the second component of
$\Lambda$ to obtain an integral manifold homeomorphic
to  $C \times (\cylinder)$ with the return map to
the cylinder factor now addition of $b(Z_0)$ (reduced mod
$1$ in the first coordinate). The lattice case is
similar, but the discrete symmetries yield integral
manifolds  $C \times (\torus)$. Depending on the
value of $b(Z_0)$, the flow on these tori can have 
$1$, $2$ or $3$ independent frequencies.

\medskip
\noindent{\bf \S 3.6 Irrational circulation ratios and quasi-periodicity.}
We now use the continuous and discrete symmetries
from a slightly different  point of view with the goal
of illuminating  the case of irrational circulation ratios and the resulting
 quasi-periodicity in the reduced Hamiltonian system.
We factor by the action of
the discrete and then the continuous symmetry. The level sets
of $J$ descend to invariant subsets, but they are no longer
level sets of a real-valued function.
The  focus here is on the lattice case
and we comment on the array
case at the  end. The nomenclature {\de $n$-torus} refers
to the topological space $\TT^n := (S^1)^n$,
\ie the $n$-dimensional manifold that
is periodic in all $n$ directions.

Recall that the phase space of the \Ham\ system
describing the vortex motions is  
$\complexes^3 - \Upsilon_\LL$,  which is denoted here as $\X$. 
The discrete action of $\integers^3$ on $\X$ factors the
phase space to $(\torus)^3-\Upsilon'$ which we give
coordinates  $(\nu_1, \nu_2, \nu_3)$, where 
$\Upsilon'$  is the factored collision set. (Note that each
$\torus$ is a Lie Group; one
does arithmetic by reducing  mod $1$ in both factors.)
The continuous symmetry of $\reals^2$  acts by simultaneous addition
on all $2$-tori, so $\tau \in \reals^2$ acts as
$(\nu_1 + \tau, \nu_2+ \tau, \nu_3+ \tau)$.
 Identifying points under this action
yields a quotient space $\hat\X = (\torus)^2-{\hat\Upsilon}$ with coordinates
$(\nu_1 - \nu_2, \nu_2 - \nu_3) := (U, V)$ and 
$\hat\Upsilon = \{ U=0\}\cup \{V = 0\} \cup \{U=V\}$. The
\Ham\ (2.4) on $\X$  descends to
$$ {\hat H}(U,V) = -{1\over 2 \pi}
( \Gamma_1 \Gt\phi(U) + \Gamma_2\Gr\phi(V)
+  \Go \Gamma_3 \phi(U + V)). \eqno{(3.11)}$$
If $\Gr$ is an integer,
the integral $J$ may be expressed in these coordinates
as $\hat J(U, V) =  U - \Gamma_3 V$ (recall that $\Go = 1$)
 with $\hat J$ having values in a 
 $2$-torus with width $\Gr$ in both directions.
If $\Gr$ is irrational, no such adaptation may be made.
But in any case, since a level set $A_0$ of $J$ is invariant
under the \Ham\ flow, the projection $\hat A_0$ to $\hat\X$ is invariant
under the \Ham\ flow induced by $\hH$.

We now treat $\hat \X$ as a subset of the $4$-torus, $\TT^4$.
If $\Gr$ is irrational, the level set $\hat A_0$ is an
immersed two-plane that winds densely in the four torus.
The  Hamiltonian $\hat H$ on $\TT^4$ restricted to  $A_0$
is  the same as the reduced Hamiltonian on $R$ given in (3.6). Thus
this Hamiltonian induces a quasiperiodic system on $R$
in the sense of [A1].
The function $\hH$ has
singularities when $U=0$, $V=0$ or $U = -V$, each of which
represents a plane in $\TT^4$. The intersection of these
planes with the densely wrapped plane $A_0$ yield the poles
of the reduced Hamiltonian $K_\LL$. Thus the collection of poles
are a quasi-crystal, at least according to one definition
of that term ([J], [A1]).

Since (the complexification of)
 $\hH$ has simple poles, the poles of the restricted
\Ham\ on any $\hAz$ will also be simple. Thus it is plausible
that the flow on
$R$ (or $\hAz$) can be treated as advection in the presence of fixed
vortices with circulations given by the residues of the poles.
The precise sense in which this is true is given  in  [SA] and [AS2].
Further, it is shown in [AS2], that this collection
of fixed vortices is dynamically fixed as well, \ie
if these vortices are allowed to interact with each other,
 they will still be stationary.
In other language, the collection of poles of the reduced
\Ham\ on $R$ forms an equilibrium configuration of
the planar $N$-vortex problem.

This viewpoint
can also be used to illuminate what happens as the various
parameters in the system are varied. For example,
once the $\Ga$ are fixed, changing $J_0$ just translates
$\hAz$ and does not alter its  ``slope''.
Thus there will be
no qualitative changes in the pattern of the poles of
$\hAz$  except for the
exceptional values of $J_0$ which
cause it to hit the  triple intersection points $U = V = 0$ (the integer
lattice).
On the other hand,  if
the $\Gamma_\alpha$'s are varied, the background flow changes as
does the slope of the $\hAz$. However, the singularity
planes  remain the same, so it is easy to track the singularities on
$\hAz$. In particular, if there is a
sequence of rational $\Gr^{(n)} = {p_n\over q_n}$ with
${p_n\over q_n} \raw \Gr^0$ with $\Gr^0$ irrational,
then the periodic systems will converge to the quasiperiodic one
after the appropriate rescalings.

A similar analysis can be applied to the array case, but
now if $\Gr$ is irrational the quasi-periodicity
of the reduced Hamiltonian $K_\AA$ will only be in
the real direction.

\medskip

\noindent{\bf \S 4 Regimes and braids.}

 Braids are the standard mathematical
tool for describing the topology of periodic motions of 
collections of points in the plane.
 In  this section braids are used to describe the  motions
of point vortices. While a braid may be used to describe the
periodic motion of any number of vortices, the analysis is
particularly simple in the case of three vortices.
As we have seen in \S3 when the total circulation is zero,
the generic motion in the reduced plane is periodic. Thus the
corresponding motion of the vortices is periodic after
adjusting for the dynamic phase, and so may be described
by a braid. 

\medskip
\noindent{\bf \S 4.1 Braids and braid types.}
This subsection gives a  brief introduction to braids from a point of
view useful for the applications that follow. For
broader perspective and additional information see
[Bm1]  or [BL]. Braids are used
to describe the motions of the bodies in the planar $N$-body problem
by Montgomery [Mt].

A {\de physical or geometric braid} is a collection of noncrossing
paths in $\reals^3$ that start at some finite 
collection of points $E$ on the plane where the third coordinate is
zero and end at the same set of points (prhaps permuted)
on the plane where the third coordinate is one.
More formally, a physical braid on
$n$ strands is a collection of  maps $\bb = 
\{b_1, \mydots, b_n\}$,  with each $b_i: [0, 1] \raw
\reals^3$ and (1) each $b_i$ has the form $b_i(t) = (a_i(t), t)$
with each $a_i$ continuous, (2) for all $t\in [0,1]$,
$b_i(t) \not = b_j(t)$ for $i\not = j$, and 
 (3) the set of initial points is
identical to the set of final points 
$E = \{ a_1(0), a_2(0), \mydots, a_n(0)\} = 
\{ a_1(1), a_2(1), \mydots, a_n(1)\}$.
The initial and final points of the braid are
collectively called the {\de  endpoints}. The collection 
of trajectories $\{ a_i(t) \}$
is called the {\de plane motion of the braid}.

A {\de mathematical braid} on $n$ strands is an element of the braid
group $B_n$. This group is defined as possessing
the  $n$ generators $\s_1, \s_2, \mydots,
\s_n$ with inverses denoted $\bs_1, \bs_2, \mydots,
\bs_n$ and the  relations $\s_i \s_j = \s_j \s_i$
for $|i-j| > 1$ and 
$\s_i\s_{i+1}\s_i = \s_{i+1}\s_i\s_{i+1}$ for all $i$ ([Bm1]).
A {\de braid word} refers to a sequence of ``letters'', with each
letter being one of the $\s_i$ or their inverses. The
inverse of a generator is indicated by an overbar, $\bs_i$. Two
braid words are said to be equivalent if they represent
the same element in the braid group, \ie one can be 
transformed into the other using the relations in the group.
The identity element in $B_n$ is denoted $e$.

The assignment of a mathematical braid to a physical one requires
a plane onto which projections are made.
The convention here is to let this plane in $\reals^3$ 
be $y=k$ for some large
negative $k$. If we  treat the $xy$ plane as the 
complex numbers, projection of \physb\ onto the chosen plane 
yields a family of curves $(\RE(a_i(t)), t)$.  An {\de ij-crossing}
is a point where  $\RE(a_i(t)) = \RE(a_j(t))$ for some $t$ and 
$i\not = j$. The  braid word will
record which strand is in front at each crossing or, more precisely, which of the 
$b_i(t)$ is closest to the projection plane.
 
The braid word corresponding to \physb\ is
read off from the picture of the projection. Assume for the
moment that all crossings are transverse and take place
at distinct times. Starting
from the bottom, examine the first crossing.
If the $i^{th}$ strand from  the left crosses behind the 
$(i+1)^{st}$, write down the letter $\sigma_i$. If
it crosses in front, write $\bs_i$. Now continue 
upward in the projection checking each crossing and writing
a braid letter. Note that the $i^{th}$ strand
at each step refers to the $i^{th}$ strand {\it from the left} at
that level. The  physical strand that is $i^{th}$ at one  level may become
$(i+1)^{st}$ or $(i-1)^{st}$  at the next level.
It is also worth noting that contrary to the conventions
of [Bm1], the convention here is that
letters further to  the right in a braid word  encode
crossings that are {\it higher} up the braid. We adopt this somewhat
nonstandard convention because our physical braids 
arise from time-parameterised trajectories in the plane, 
and it is more natural to visualise these in $\reals^3$ with
time going upwards.

It is  natural to ask  how much the mathematical braid
tells us about the physical braid. In particular, which physical
braids get assigned the same mathematical one?
Say that two physical braids $\bb$ and $\bb'$ 
with the same endpoints are {\de equivalent} if
one can be obtained from the other by a deformation that 
fixes the endpoints  and does
not cut the strands, \ie there is
a continuous family of physical braids $\bb^s$ for $s\in [0,1]$
with $\bb^0 = \bb$ and $\bb^1 = \bb'$.  
The relations in the braid group are chosen precisely so that 
two  physical braids are
 equivalent exactly when they are assigned equivalent braid words,
 \ie the same mathematical braid. 
A theorem of Artin's ([Bm1]) says that 
any equivalence between physical braids can be
described in terms of just the two kinds of deformations
described by the relations in the braid group. 

Because equivalent physical braids are assigned equivalent
braid words we can  eliminate the assumption
we had to make in order to assign a mathematical braid to a physical one.
 If \physb\ does not have transverse and time-distinct crossings,
deform it to another braid $\bb'$ that does and 
compute a mathematical braid for $\bb'$. Since any
other ``good'' deformation will be assigned a braid 
word equivalent to that of $\bb'$, this braid word may
be used unambiguously for \physb.

Thus far we have restricted attention to equivalence
of physical braids with the same endpoints. Since the
braids here arise from vortex motions with a variety 
of initial positions we need to extend the notion of equivalence.
Informally two braids with perhaps different endpoints
are equivalent if one can be transformed into the other via a plane
transformation applied on each horizontal plane followed
by a  deformation with fixed endpoints.
More formally,  two physical $n$-braids 
\physb\ and $\bb'$ are {\de equivalent} if there is a \homeo\
$h:\reals^2\raw \reals^2$ with $h(a_i(0)) = a_i'(0)$ for all
$i$ and the physical braid $\{ (h(a_i(t)), t)\}$
is equivalent with fixed endpoints to $\bb'$.
A simple argument shows that physical braids that are equivalent
in this sense are assigned conjugate elements of
the braid group, \ie the words satisfy $w' = g w g\inv$,  with
$g$ an element of the braid group that  represents
the \homeo\ $h$. 
Note that changing the projection plane may be accomplished  by
a rotation in the plane, and so the mathematical braid  obtained using the
new projection plane will be conjugate to the original.

Since the goal here is to use braids to analyse
the  topology of vortex
motions, topologically equivalent physical braids and 
physical braids viewed by different observers must
be assigned the same mathematical object.
The remarks of the  last paragraph make it clear that this object 
cannot be just a mathematical braid, but rather  must
be a conjugacy class in the braid group, \ie the collection of
all elements conjugate to a given one (and thus conjugate to
each other). These conjugacy classes 
were called {\de braid types} in a related context and that terminology is adopted
here (cf. [Bd]). 
For  simplicity of exposition in the sequel  we will often speak
informally of the braid associated to a physical braid, but
a more careful terminology  would be braid type.

\medskip
\noindent{\bf \S 4.2 Three vortices in the plane.}
In assigning braids to  the motion of three point vortices in the
infinite plane we continue to focus on
the case of zero net circulation, $\sum \Ga = 0$. As in \S3, we
fix a value of the integral $J_0 = Q_0 + iP_0$,
perform the reduction, and obtain
the Hamiltonian (3.6) on a copy of the plane $R$ with complex
coordinate $Z$. 

Pick an initial condition $Z_0$ in $R$ which is
contained in a \po\  $Z(t) := Z(t; Z_0)$ with period $P$.
Using (3.4), the motion $Z(t)$ determines the evolution of
the differences in the positions of the three  vortices. 
Thus $Z(t)$  can be used  to compute the
motion   in the frame of one of the vortices.
In the $z_2$-frame  this motion is described by
$$\eqalign{ Z_1(t) &:= z_1(t) - z_2(t)= Z(t)\cr
 Z_2(t) &:= z_2(t) - z_2(t) \equiv 0 \cr 
Z_3(t) &:= z_3(t) - z_2(t) = {J_0 - \Gamma_1 Z(t) \over \Gamma_3}.\cr}
\eqno{(4.1)}$$
Thus all the $Z_\alpha$'s are periodic with period $P$.
Note  that $\za \not = \zb$ for 
$\alpha \not = \beta$ (\ie the absence of collisions between vortices) 
is equivalent to  $\Za \not = \Zb$ for $\alpha \not = \beta$.
Again using (3.4), this happens as long as $Z(t)$ avoids the points
$p_1 := J_0/\Gamma_1, p_2 := - J_0 / \Gamma_2$, and
$ p_3 := 0$. The points $p_\alpha$  are the {\de poles} of the 
Hamiltonian $K_\PP$ from (3.6).

The vortex motions generate a physical braid by 
treating the vertical direction as time and
defining 
$$\hZ_\alpha = (\Za(t), t),\eqno{(4.2)}
$$
for $\alpha = 1, 2, 3$. Each path $\hZ_\alpha$ then connects
a point on the plane $t = 0$ to the same point on the plane 
$t=P$, and further, since $Z(t)$ does not pass through any of the poles,
the paths do not intersect.  Thus 
the paths yield a physical braid on three
strands, which may be assigned an element of $B_3$ as in
the previous subsection.

This assignment is most easily accomplished by taking
a more topological point of view. If we 
define $\Ro = R -\{p_1, p_2, p_3\}$, then the  \po\ $Z(t)$ in
$R$  represents a closed curve (or loop) in $\Ro$. The procedure
 used to construct a physical   braid
from  $Z(t)$ can equally well be used to assign one 
to any loop in $\Ro$. If 
 $\gamma$ is a loop in $\Ro$, \ie  $\gamma:[0,1]
\raw \Ro$ with $\gamma(0) = \gamma(1)$, let  
$Z_1, Z_2, Z_3$ represent the three paths
given by (4.1) using $Z(t) = \gamma(t)$. The physical braid
corresponding to $\gamma$ is then generated by (4.2).

The computation of  the mathematical braid corresponding to
 this physical braid just requires  knowledge of 
the crossing of its strands.
Since we are viewing from  the negative imaginary axis,
the positions of the $Z_\alpha$ from left to right are determined
by their real parts. Strands cross exactly
when this  order  changes. This  can only happen when
$\RE(\Za(t)) = \RE(Z_{\alpha'}(t))$, or equivalently,
 when $\RE(Z(t)) = \RE(p_\beta)$ 
where $\beta $ is the index different from $\alpha$ and
$\alpha'$. Thus the vertical lines through
the poles  divide $\Ro$ into vertical strips
in which the order of the $Z_\alpha$ from left to right is constant. 
These vertical lines will be called {\de crossing lines}.
  
To specify the generator corresponding to the crossing of
one of  these lines it is also
necessary to know  which strand is in front.
This information is determined by the sign of $\IM(Z(t_c))- \IM(p_\beta)$ 
when $Z(t_c)$ lies on the  crossing line $\RE(Z) = \RE(p_\beta)$.
This sign can only
change on a crossing line when $\IM(Z(t_c))- \IM(p_\beta) = 0$,
\ie at the pole.  Thus all the crossings on the same side
of the pole correspond to the same strand being in front. 
We will call a component of a crossing line minus its pole a {\de generator  arc}. 
The final piece  of information needed to specify the
generator describing a crossing is the direction that 
$\gamma$ traverses the generator arc. In summary then, any
two crossings of a generator arc in the same direction  
contribute the same generator to the braid. Thus to compute
braids of loops or actual vortex motions it suffices to 
compute the generators corresponding to each generator arc. 
A sample computation is given in the next subsection.

There are two issues that need to be clarified before moving
on to the example.
First,  unlike a mathematical
loop there is no distinguished starting point for a
periodic motion of the vortices. However, changing the initial point 
on $ Z(t)$  will only cyclically permute the generators in the
braid word representing the motion. This corresponds to conjugating 
the word in the braid group and this ambiguity has
already been resolved using the braid type.
The second issue is that the usual generators
of the braid group $B_3$ only keep track of 
which adjacent strands are crossing and not the 
numbering of the stands.  Thus in assigning a braid 
generator to a crossing we have lost track of which vortex has
crossed which; our braid word is only encoding
the topological type of the interaction, and not
whether, say, vortices 2 and 3 are the pair 
that are circling each other.
However this information can be recovered easily by 
knowing the relative positions of the vortices at the
starting point of the braid and following the strands. 

\medskip

\noindent{\bf \S4.3 A planar example.}
Figure 4.1 shows the crossing lines for 
 the planar case of Figure 3.1. The pair of numbers at the
top of the crossing line indicate which pair of vortices are crossing.
 Each generator arc has a number $j$ and an arrow;
crossing the arc in the direction of the arrow contributes
the positive generator $\sigma_j$ to the braid word. Crossing in the opposite
direction contributes  the inverse of this generator $\bs_j$.
The braid word corresponding to a loop $\gamma(t)$ can be computed 
by writing in order the generators corresponding
to $\gamma (t)$'s  sequential crossings of generator arcs.
\midinsert
\def\epsfsize#1#2{.4\hsize}
\centerline{\epsfbox{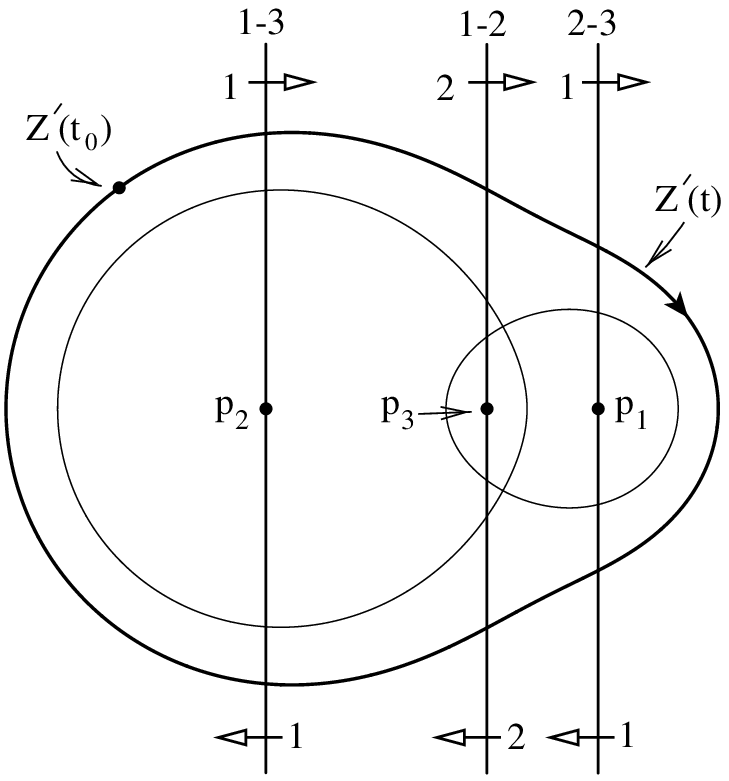}}
\medskip
{\leftskip=60pt\rightskip=60pt\parindent=0pt

{\bf Figure 4.1:}
The crossing lines and generator arcs for the system
shown in Figure 3.1.

}
\endinsert

As an example, let us compute the braid word corresponding to 
\po\ $Z'(t)$ indicated in the region labelled D in Figure 4.1. 
 Begin tracking the motion
at the starting point labelled $Z'(t_0)$  in the upper left hand corner.
At this point the vortices
from left to right are $1, 3, 2$. The first generator arc crossed 
is above the pole $p_2$, it contributes a $\s_1$ to the braid word. Moving
across horizontally, the next generator arc contributes a $\s_2$,
and so on. When we return to the initial point the entire braid 
word $\so \st \so\so\st\so$  has been read off. This braid is shown
in Figure 4.2c, with the strands labelled below. 
By examining the braid we see that this motion corresponds to the vortices
rotating clockwise around each other. Examining the braid one
can see that the motion is topologically the same as all
the vortices rotating once around a circle clockwise.
This becomes especially clear after  using  the  braid relation 
and rewriting the braid as $(\so\st)^3$.
\midinsert
\def\epsfsize#1#2{.7\hsize}
\centerline{\epsfbox{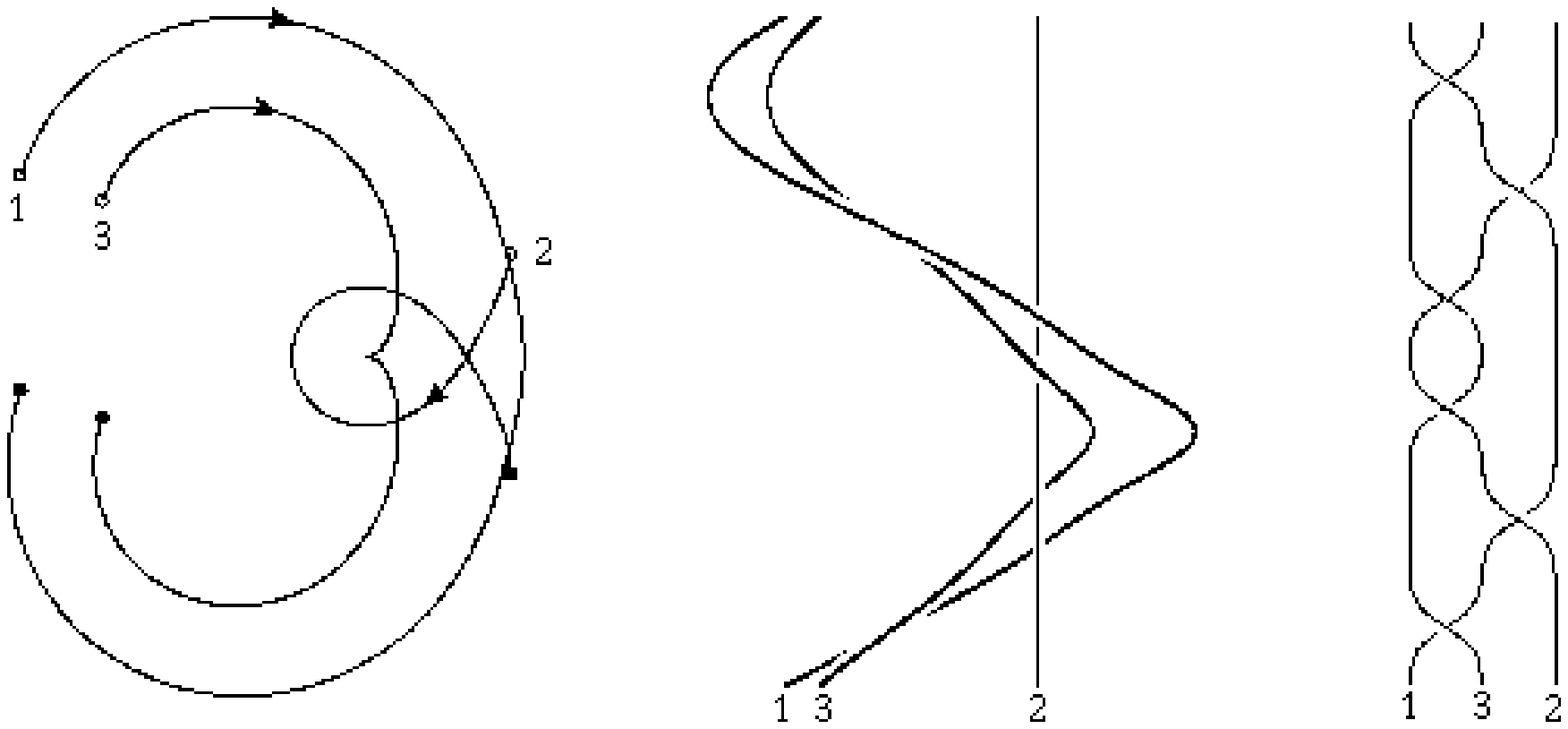}}
\medskip
{\leftskip=60pt\rightskip=60pt\parindent=0pt

{\bf Figure 4.2:}
Vortex motions from regime D of Figure 3.1. (a) Trajectories of
the three vortices in the plane. (b) Physical braid of this motion
in the frame of the second vortex. (c) Mathematical braid of the motion.

}
\endinsert

Figure 4.2a shows the planar trajectories of 
three vortices corresponding to the \po\ $Z'(t)$.
Recall from \S 3.2 that there are many vortex motions
corresponding to  a given  \po\  in  $R$, but
all these motions differ by a uniform translation, so it suffices
to pick an initial position of one of the vortices, say $z_2(0) = 0$.
Figure 4.2b shows the physical braid corresponding to the
vortex motion in the  frame of the second vortex
as given by (4.1) and (4.2).

The  other regions  A, B, and C each contain a pole. 
The braid describing each of these regimes is the square
of a generator  and  corresponds to a pair of vortices
rotating  about each other once per period. 
There is no linking with the third vortex. The pair
of vortices that are involved in the
rotation depends on the  pole in question, and the direction
of the rotation (\ie\ whether the braid is $\s_i^2$ or
$\bs_i^2$) depends on the signs of the circulations of the interacting
vortices.

Recall that a {\de separatrix} is an orbit connecting two 
saddle points. Define a {\de regime} as
a  connected component  of $R -\{\hbox{poles\ and\ separatrices} \}$.
It is clear from Figure 4.1  that every \po\ in the same regime 
yields the same sequence of crossings of generator arcs, and
thus the same braid word. Therefore all the motions within a regime have
the same topological type of motion.
More generally,  any pair of loops that 
 can be continuously deformed into each other
(\ie are homotopic) in $\Ro$  will
be assigned the same braid; homotopic loops may cross different
generator arcs, but any extra crossings will consist of
a collection of crossings and then reverse crossings of the
same generator arcs. From another point of view, deforming the loop
corresponds to deforming the physical braid. The corresponding
mathematical braid will only change when a pair of strands go
though each other. This can only happen   when the
loop passes through a pole, which is not allowed  as all our
loop deformations are in  the complement of the poles.

This situation is somewhat analogous to the residue
theorem in which a deformation of a closed path in the complement of
the poles does not change the value of the integral. However, the situation
here is more restrictive; {\it homologous}  loops yield
the same integral but  only {\it homotopic} loops give
the same braid.  In algebraic  language,
after fixing a basepoint for loops, the process described here 
gives a homomorphism from
the fundamental group of the punctured plane, $\pi_1(\Ro)$,
to the braid group on three strands, $B_3$.

Although all \po s in the same regime have the same braid,
it is important to note that the period $P$ and dynamic
phase $b$ of the \pos\ in a single regime can vary greatly. 
The period will go to zero near a pole and approach infinity 
as orbits near a separatrix. Thus the time scale of the actual vortex 
motions in the same regimes can be very different. In addition, 
variations in the dynamic phase can make  
a significant difference in the motion as 
observed in the lab frame.

\medskip
\noindent{\bf \S 4.4 Three-vortex arrays.}
This subsection  develops the tools needed to
assign a braid  to the periodic motion of 
three vortices in a singly-periodic array. 
We continue to restrict to the case of zero net circulation $\sum \Ga = 0$,
fix a value of the integral $J_0 = Q_0 + iP_0$, and
perform the reduction as in \S 3. The result is 
a Hamiltonian system given by (3.6) on a copy of the plane $R$ 
with complex coordinate $Z$. If $\Gamma_3$ is the rational $p/q$ (in
lowest terms),
 then  the system on $R$ has a mod $p$ symmetry in the real part.

By treating each vortex position $\za$ as a singly-periodic
variable, the vortex motion can be viewed as taking
place on a cylinder $\circle\times\reals$ (the circle  $\circle$ here
has perimeter one since  we  continue to restrict to 
the case $L=1$).  Equivalently, in the language of \S 3, we 
factor out all the motions by the discrete symmetry. Again
we eliminate the dynamic phase by passing
to the frame of the second
vortex, thus making the motion periodic. Accordingly, let 
$$
\eqalign{ c_1(t) &:= z_1(t) - z_2(t)= Z(t)\cr
 c_2(t) &:= z_2(t) - z_2(t) \equiv 0 \cr
c_3(t) &:= z_3(t) - z_2(t) = {J_0 - \Gamma_1 Z(t) \over \Gamma_3},\cr}
\eqno{(4.3)}
$$
in which properly speaking the subtraction is done in the Lie group
$\circle\times\reals$. The triple $ (c_1(t), c_2(t), c_3(t))$ is called
the {\de cylinder motion} of the vortices.

A braid describes the motion of a set of points in the
plane. To translate the cylinder motion
 to the plane, recall that the cylinder is 
topologically equivalent to the punctured complex plane $\complexes -\{  0 \}$.
 This topological equivalence is realized  by the conformal map 
$T(z) = \exp(2 \pi i z)$. Under this conformal map a
path around the cylinder transforms to  a path 
 around the origin in the complex plane.
To capture this type of motion in the braid description, we
add the constant path at $0$ as the 
 last coordinate of the motion  in the plane.
Thus the  {\de plane motion} of the  vortices is given by 
$$ 
\eqalign{ s_1(t) &:= T(c_1(t)) = \exp(2 \pi i Z(t))\cr
 s_2(t) &:= T(c_2(t)) \equiv  1 \cr
s_3(t) &:=T(c_2(t)) = \exp(2\pi i{J_0 - \Gamma_1 Z(t) \over \Gamma_3})\cr
s_4(t) &\equiv 0.  \ \cr}
\eqno{(4.4)}
$$

Using the same technique as in the  previous subsection we associate a
braid (now on $4$ strands) to the plane motion 
of a three-vortex array. Again we visualise the plane motion in
three dimensions using (4.2) with time going upward and view the 
resulting strands from the negative imaginary axis. 
The {\de crossing curves} in  $R$  describe the
positions of an orbit $Z(t)$ at which $s_\alpha$ and $s_\beta$
cross; this happens when 
their real parts are equal. To express this in equation form,
define four functions on $R$ by $\psi_1(Z) = T(Z), \psi_2(Z) = 1,
\psi_3(Z) = T((J_0 - \Gamma_1 Z)/\Gamma_3),$ and $\psi_4(Z) = 0$.
Then $s_\alpha(t) = \psi_\alpha(Z(t)) = T(c_\alpha(t))$, and
the $(\alpha,\beta)$ crossing lines are defined by the
equation $\RE(\psi_\alpha(Z)) = \RE(\psi_\beta(Z))$.
In contrast to the planar case, these equations no longer 
yield just straight lines and thus the change in terminology
to crossing {\it curves}.

The strand which is in front at a crossing is determined by the sign of
$\IM(s_\alpha(t_c))- \IM(s_\beta(t_c))$ when $\gamma(t_c)$ is on
the corresponding crossing curve. This sign changes
on the crossing curve exactly when $s_\alpha = s_\beta$, which corresponds
to the poles of the Hamiltonian (3.6). Note  that it is no longer the
case that every crossing curve contains a
pole. This means that  crossing anywhere on the curve
yields the same generator of the braid word, \ie the entire curve
is a single generator arc.
 
\medskip
\noindent\noindent{\bf \S 4.5 An  array example.} We now perform the calculations
described in \S4.4 on the  array
case shown in Figure 3.2. See [SA] for a details on this
and other array examples. 
The crossing curves corresponding to
$s_1$ and $s_3$ are defined by $\RE(\psi_1(Z)) = \RE(\psi_3(Z))$, 
which, writing $Z = x + iy$,  is 
$$\exp(- 2 \pi y) \cos(2 \pi x)  = 
\exp(- 2 \pi ({2\over 3} y - {1\over 4}))
\cos(2 \pi ({2\over 3} x - {1\over 12})).
$$
Both cosine terms can vanish yielding the
 two vertical lines $x = 5/4$ and
$x = 11/4$.  Otherwise we may solve for
$$y = {3\over 2 \pi}
(\log({\cos(2 \pi x) \over \cos(2 \pi ({2\over 3} x - {1\over 12}))})
- {\pi/2})
$$
which is defined only on the (mod 3) intervals $(-3/4, 1/4), (1/2, 3/4)$, and
$(7/4, 2)$. Thus there are five distinct $(1,3)$-crossing curves.
The computations for other pairs is similar and  the  various crossing
curves are shown in Figure 4.3a. The numbers separated by a hyphen
indicate which
pair of $s_\alpha$'s are crossing. Note that since the vertical lines 
$x = 5/4$ and $x = 11/4$ correspond to  $0 = \RE(s_1) = \RE(s_3)$, and
we are considering $0$ as the position of $s_4$, these vertical lines
correspond to a triple crossing.
\midinsert
\def\epsfsize#1#2{.7\hsize}
\centerline{\epsfbox{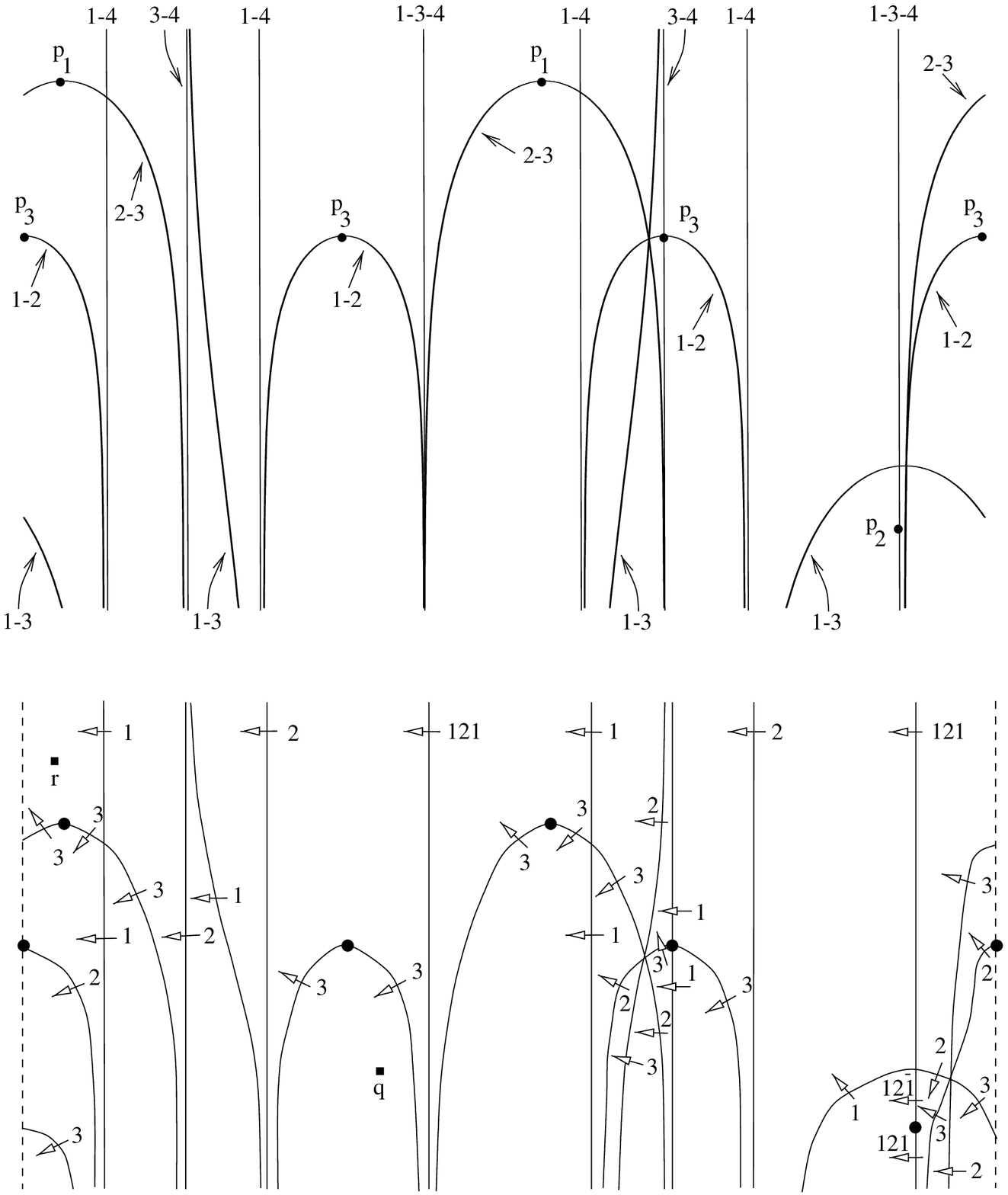}}
\medskip
{\leftskip=60pt\rightskip=60pt\parindent=0pt

{\bf Figure 4.3:}
(a) Crossing curves  and  (b) generator arcs  for the
system in Figure 3.2 (the picture in (b) has been
distorted for increased clarity). The points labelled
$q$ and $r$ correspond to initial points of trajectories
as explained in the text.

}
\endinsert

The poles are the points $0, 1$ and $2$ corresponding to $s_1 = s_2$,
 the points ${1\over 8} + {3\over 8}i$ and   ${13\over 8} + {3\over 8}i$
 corresponding to $s_2 = s_3$, and the point 
  $-{1\over 4} - {3\over 4}i$ corresponding to $s_1 = s_3$.
Note  that many crossing curves do not contain a
pole. Figure 4.3b shows the results of computing
the generator arcs, with labels using the
same conventions as in Figure 4.1. 

It is important to note  that the details of Figure 4.3 are not
intrinsic to the vortex motions. Changing the projection plane
used to compute the braids would change all the crossing
curves. However, the braid which is computed for a loop is
intrinsic up to conjugacy.
In algebraic language, after fixing a basepoint, the equations
(4.4) (using $Z(t)$ to represent a general loop)
may be  used to define a homomorphism of
 $\pi_1(R - \{ \hbox{\rm poles}\}) \raw
B_4$. This homomorphism is not intrinsic, but it is intrinsic up
to conjugacy in $B_4$.

\tenpoint
\topinsert\centerline{
\vbox{\offinterlineskip
\hrule
\halign{&\vrule#&\strut\myquad\hfil#\hfil\myquad\cr
height3pt&\omit&&\omit&&\omit&&\omit&\cr
&Regime && Braid  && TN type  && Expansion &\cr
&\omit && \omit  && \omit  && Constant &\cr
height2pt&\omit&&\omit&&\omit&&\omit&\cr
\noalign{\hrule}
height2pt&\omit&&\omit&&\omit&&\omit&\cr
&I && $(\bso\bst)^3 \bso^2(\bst \bso)^3$ &&  reducible, all f.o. &&\omit&\cr
&II && $\sr\st  (\so\st\sr)^2 \st\sr \so\st\so  (\so\st\sr)^2
\st\so\sr  \so\st\so
\sr $ &&  reducible, all f.o.  &&\omit&\cr
&III &&$\sr^2 $&&  reducible, all f.o.  && \omit&\cr
&IV &&$\sr^2$ &&   reducible, all f.o. && \omit&\cr
&V &&$\so^2$ &&   reducible, all f.o.  && \omit&\cr
&VI &&$\bst^2 $&&  reducible, all f.o.  && \omit&\cr
&VII &&$\sr^2$ &&   reducible, all f.o.  && \omit&\cr
&VIII &&$\bsr^2 $&&   reducible, all f.o.  && \omit&\cr
&IX &&  $\sr^2\st^2$ &&   reducible, all f.o.  && \omit&\cr
&X && $(\sr\st)^3 $&&   reducible, all f.o.  && \omit&\cr
&XI&& $(\sr\st)^2 \sr \so \st\so^2  \st  \so  \st \sr \so \sr
 (\bso \bst)^3\bso $
  &&  pA
&& 13.93&\cr
&XII && $(\sr\st)^2 \sr \so \st\so^2  (\st \sr \so \sr )^2
 (\bso \bst)^3\bso $
 &&   reducible, one  pA  && 13.93&\cr
&XIII && $(\bso \bst)^2  \bsr^2 (\bst\bso)^5$&&  pA
   && 9.90&\cr
height2pt&\omit&&\omit&&\omit&&\omit&\cr}
\hrule}
}
\medskip\twelvepoint
{\leftskip=60pt\rightskip=60pt\parindent=0pt

{\bf Table 4.1:}
The mathematical braids and \TN\ type of the regimes
for the system shown in Figure 3.2. The expansion constants
for pseudoAnosov (pA) components are given. 
In the finite order (f.o.) case these constants are $1$. This means 
that there is  no intrinsic topological expansion in that case.

}

\endinsert

\twelvepoint

Now define a regime as in the planar case, \ie as a
connected component  of 
$R -\{\hbox{poles\ and\ separatrices} \}$.
Each \po\ in a regime gives rise to the same mathematical braid;  these
are listed in the second column of Table 4.1. Note that the relations
in the braid group have been used to simplify many of the braid
words.  The braids for regimes I and IX through XIII  were computed
using closed trajectories with initial position near the point labelled
$r$ in Figure 4.3b. The initial position used for regime II was
near the point labelled $q$.

As a sample we compute the mathematical braids
corresponding to regimes I and  XIII.
The closed orbits in regime I travel across
the top of the reduced plane from left to right.
To compute the 
braid we chose the point labeled 
$r$ in Figure 4.3b as the initial point. Moving to the right from this point
we first traverse a vertical line on which there is an arrow labeled $1$. 
This indicates that the motion of the trajectory
has resulted in a crossing of the two left most strands in the
physical braid representing the vortices. Further, since 
the trajectory's motion is opposite to
that of the arrow, the strands are crossing with the left one in
front of the right one. Thus the traversing of the first 
crossing curve contributes the letter $\bar \sigma_1$ to the
braid word of the trajectory. Continuing to the right the trajectory
traverses a crossing curve on which there us an
arrow labeled $2$ and the arrow points 
in the direction opposite to the traverse. Thus this motion
contributes a $\bar \sigma_2$ to the braid word of the trajectory.
Continuing across the top of Figure 4.3b, we add braid letters
$\bar \sigma_1$, then  $\bar \sigma_2$, etc.  After encountering the
right edge  of the phase plane the identification with the
left edge allows a return to the initial point.
The entire braid word for this regime is thus 
$\bso \bst\bso\bst\bso\bst\bso\bso\bst\bso\bst\bso\bst\bso =
(\bso\bst)^3 \bso^2(\bst \bso)^3$.

The closed trajectories for regime XIII have a motion similar
to those  of regime I 
with the crucial topological difference that they
pass below the pole  $p_3$ which is located at  $1 + 0 i$. This implies that
in the complement of the poles, the closed loops corresponding to
regimes I and XIII are {\it not} homotopic. Beginning a closed trajectory 
representing regime XIII near the point 
$r$, the braid word starts with $\bso \bst\bso\bst$
like that of regime I, but since the loops for regime XIII pass below
the point $p_3$, their braid words have a $\bsr\bsr$ next, and then 
continue with the same letters as regime I. The entire
braid word for regime XII is thus 
$\bso \bst\bso\bst \bsr\bsr \bso\bst\bso\bso\bst\bso\bst\bso\bst\bso$.
Note that the braid relation $\bso\bst\bso =
\bst\bso\bst$ used on the triple of letters right after
the $\bsr\bsr$ allows us to reduce the word to
$(\bso \bst)^2  \bsr^2 (\bst\bso)^5$ as given in Table 4.1.

We now briefly describe some of the
motions. In these descriptions we will use the word ``vortex''
to refer to the trajectories of the planar motions, $s_1, s_2, s_3$, 
and $s_4$. Properly speaking $s_1, s_2$, and $s_3$ are the
transformed positions of the vortices, and $s_4$ is not a vortex
at all, but rather it keeps track of the position of the origin.

Before interpreting the braid, it is useful to recall
the various transformations that have been used. 
The motion in the singly periodic plane is really
of infinite families of vortices. These families are
treated as three individual vortices on a  cylinder 
and then this motion is 
transformed to the punctured plane. Thus a strand in
the braid rotating about 
 the  strand representing vortex 4 ($=$ the origin)  describes motion
around the cylinder, which in turn, is  a horizontal  translation by one 
period of the corresponding family of vortices in the singly-periodic
plane. In addition, the braid is obtained 
using the frame of the second vortex, so 
in particular vortex 2  will never rotate around the origin,
and thus the strands corresponding to vortex 2 and 4
 will always be parallel.

The  regimes with the simplest braids are those that
contain a pole, III through VIII. 
The vortex motions are topologically the same as regions A, B, and C
in the plane case: a pair of vortices are circling each 
other and the other vortex is uninvolved. There is no rotation
around the origin, thus there
is no net rotation around the cylinder. 

The vortex motions corresponding to 
 regime  X also have no net  motion around the 
cylinder (see figure 6.1a). This is indicated by the fact  
that none of the other strands 
link with the far left  one which represents the origin of $\complexes$.
Examining the right hand sub-braid on 3 strands, one sees that
every  vortex rotates about every other one
once in the clockwise direction. The net motion is the same as that
in the planar region D and it amounts to one full rotation
of the vortices as a group. The motions in regime IX are similar,
but the sub-braid is slightly different;
 vortices 2 and 3 rotate about
each other clockwise, and then vortices 1 and  3 do likewise, but
vortices 1 and 2 do not link.

The  regimes with the next simplest braids are I and II.
 The mathematical braid for I is shown in Figure 4.4d. 
 Vortex 1 rotates three times around
the origin counter clockwise and vortex 3
rotates twice, also counter clockwise. This rotation takes place ``inside'' 
the fixed vortex 2. In the singly periodic plane, this corresponds
to the vortex 1 family translating three strips to the right
and the vortex 3 family two strips;  this translation takes
place below the position of vortex 2. This motion 
is shown in Figure 4.4a with one vortex  from each singly periodic
 family pictured. Figure 4.4b shows  the
motion in the frame of the second vortex after it
has been transformed to the plane.
The physical braid of this motion constructed using 
 (4.4) and (4.2) is shown in  Figure 4.4c.
The motion in regime 
II is similar, but now  the rotation is clockwise, and goes around
vortex  2, which simply means that in the singly-periodic plane
the translations of vortex 1 and 3 are above the position of
vortex 2.

\midinsert
\def\epsfsize#1#2{.8\hsize}
\centerline{\epsfbox{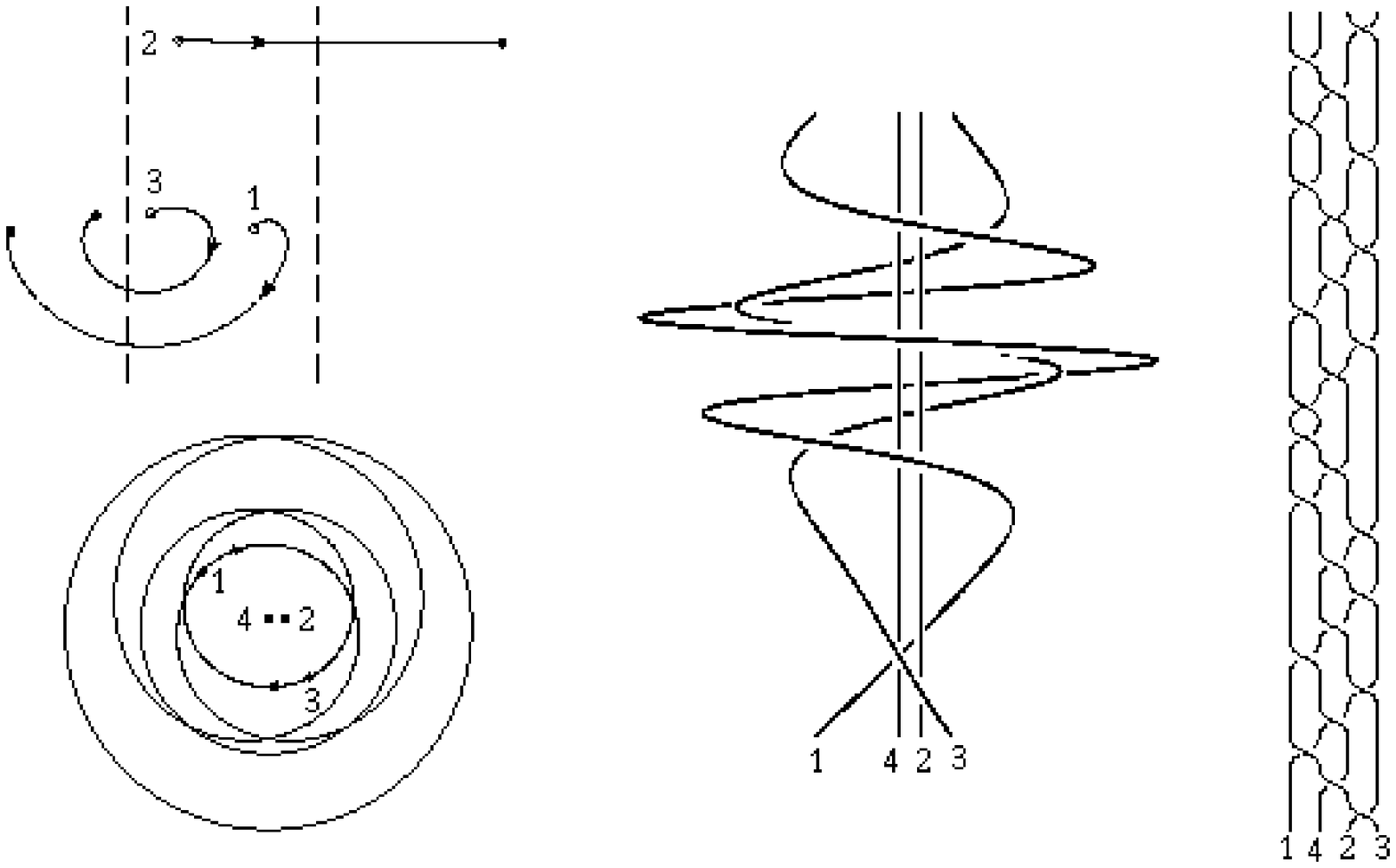}}
\medskip
{\leftskip=60pt\rightskip=60pt\parindent=0pt

{\bf Figure 4.4:}
Vortex motion from regime II of Figure 3.2. (a) Trajectories
of the three vortices in the singly-periodic plane (only one
representative of each family is shown).
(b) The motion in the frame of the second vortex transformed into the plane.
(c) Physical braid of the transformed motion.
(d) Mathematical braid of the transformed motion.

}
\endinsert

The motion in regimes XI, XII, and XIII is sufficiently complicated
that the braid itself is  the best description of
the motion. One feature of interest in   regime XII (see Figure 6.1b) is
that the vortex pair 1 and 3 have the same motion
with respect to the other two strands  while they rotate
about each other three times. This kind of hierarchy of motion
is  described precisely by the notion of reducibility introduced 
 in \S5.

\medskip
\noindent{\bf \S 4.6 Three-vortex lattices.}
The method
of analysis of this section can also be applied to vortex lattices.
When $\Gamma_3=p/q$ is
rational, the reduced Hamiltonian system (3.6) can be viewed
as taking place on a torus with width $p$ in both directions. The
generic orbit of this system will be a periodic orbit $Z(t)$. This
\po\ can be used to generate a motion of the three vortices
on the torus which will be periodic in the frame of one of the vortices.
However, in contrast to the array case, this cannot be transformed
into a motion on the plane, and so one must record
the motion using the braid group of the torus. This group
not only has generators corresponding to the switching of two
points as in the usual braid group, but  it also has generators
describing motions that circulate around the meridian and longitude of
the torus (see [Bm2]). Thus the braids are more difficult
to visualise. However the basic features of the analysis in the array and
plane case go through with no difficulty. In particular,
each regime is assigned just one braid, and so the braid captures the
sense in which all the vortex motions in one regime are topologically the
same and distinct from those of other regimes.

\medskip

\noindent{\bf \S 5 Isotopy classes and the \TN\ theory.}

This section provides an introduction to
isotopies and the \TN\ theory in preparation for the study of the 
advection \homeos\ generated by  vortex motions. Roughly
speaking, two \homeos\ are isotopic if one is a
continuous deformation of the other.
The braid of a vortex motion will
be seen to determine the isotopy class of the corresponding
advection \homeo.  The  braid is then used as input into
the \TN\ theory which describes dynamical data that is shared by 
isotopic \homeos, and  is  thus present in the advection \homeo.

\medskip
\noindent{\bf \S5.1 Homeomorphisms, isotopy, and braids}. 
 A one-to-one and  onto map $f : X \raw X$ is called a {\de \homeo\ } if 
 $f$  and its inverse, $f\inv$, are both continuous. If in addition,
$f$ and $f\inv$ are both differentiable, then $f$ is called a {\de \diffeo}. 
Two \homeos\ $f$ and $g$ are {\de conjugate} if there is a third
\homeo\ $h$ with $f = h g h\inv $. The \homeo\ $h$ represents the
change of coordinates from $f$ to $g$. If $E$ is a finite set and
$f(E) = E$, then $f$ is called a {\de rel} $E$ {\de \diffeo } or
{\de \homeo}.

Most of the  \homeos\ occuring in this paper 
are defined using the  advection of a passive particle 
in a time varying velocity field. Given a vector
field on the plane $(u,v)$, the advection equations are 
$$ {dx\over dt} = u(x, y, t)\ \ \  {dy\over dt} = v(x, y, t).
\eqno{(5.1)}$$
Letting $z = (x,y)$, the solution with position  $z_0$ at time $t_0$ 
is denoted  $z(t;z_0, t_0)$. By convention the 
running time $t$ satisfies $z(t_0; z_0, t_0) = z_0$, {\it not }
$z(0; z_0, t_0) = z_0$ (unless, of course, $t_0 = 0$). 
Fixing a time $t$, the {\de time} $t$-{\de flow map} or the {\de time}
 $t$-{\de advection
\homeo\ } is defined as $h_t(z_0) := z(t;z_0, 0)$, \ie we advect
each particle for time $t$, and $h_t$ maps each initial position
 to its final position.
The collection of $h_t$ for all $t$ describes all the solutions and is called a 
{\de fluid motion} (cf. \S3.1 on the use of ``flow''). Note that given a fluid
motion the underlying velocity field is  obtained by differentiation.

The usual definition of isotopy rel a finite set $E$ is the following.
Given two rel $E$  \homeos\ $f_i:\reals^2\raw\reals^2$, 
say that $f_0$ is {\de isotopic to} $f_1$ {\de rel} $E$ if they 
can be connected by a continuous family of
 rel $E$ \homeos\ $f_t$, $t\in [0,1]$.
Since $E$ is a finite set this implies that 
 $f_t(k) = f_0(k)$ for all $t$ and $k\in E$.  The collection
 of $\homeos$ isotopic to $f$ rel $E$ is called its
 {\de rel $E$ isotopy class}.
The finite set $E$ in the sequel will be the initial positions of the
collection of point vortices. These points will be fixed under 
the advection \homeo\ considered, but in general, the points
of $E$ could be permuted by a rel $E$ \homeo. It is sometimes convenient
to remove the set $E$ from the plane and consider a rel $E$ \homeo\
as a \homeo\ of the punctured plane $\reals^2 - E$. Note that
two  rel $E$ \homeos\ are isotopic rel $E$ exactly when they are
isotopic as \homeos\ of $\reals^2 - E$. 

The exposition of isotopy in fluid mechanical terms is most
clear in the special case of isotopy to the identity rel $E$.
For simplicity we restrict to the case of \diffeos\ as 
is natural for fluid mechanical applications. 
A rel $E$ \diffeo\  $f:\reals^2\raw\reals^2$ will be 
 isotopic to the identity rel $E$ exactly when $f$ is the time 
$1$ advection \homeo\ of a velocity field which satisfies
$u(k, t) = v(k,t) = 0$
for all $k\in E$ and $t\in [0,1]$. This means that all the points of
$E$ are fixed throughout the advection. Thus we may think of $f$ as arising
from advection due to body forces on the fluid and the points
in $E$ as fixed point-like regions that are not permeable
by the fluid. If the points of $E$ move under a given fluid motion,
the time-one flow map can  still be isotopic to the identity if
there is {\it another}  fluid motion that also gives rise to $f$, but
in this new motion the points of $E$ are fixed for the entire motion.

The braid associated with the motion of
points in $E$ can be used to determine whether a given
advection \diffeo\ is isotopic to the identity. Given a fluid
motion $h_t$, the motion of the points in $E$ is given
by  $a_i(t) := h_t(k_i)$ for each $k_i\in E$. Now
a family of \diffeos\ $\psi_t:\reals^2\raw\reals^2$ for $t\in [0,1]$
 with $\psi_t(a_i(t)) = k_i$ for all $k_i\in E$ and
$\psi_0=\psi_1 = id$  induces a level preserving
\diffeo\ $\Psi:\reals^2\times[0,1]\raw\reals^2\times[0,1]$  
defined by $\Psi(z, t) = (\psi_t(z), t)$. The \diffeo\ $\Psi$ 
 will take $\bb$ to the trivial physical braid which
consists of just vertical strands at the points
of $E$. If this is the case say that $\bb$ is
\wonko\ to the trivial braid $\trivb$.

The relationship of braids to isotopy classes is
this: given a   fluid motion $h_t$ with advection
\homeo\ $f=h_1$ and  a finite set $E$ with $f(E)= E$, then
$f$ is isotopic to the identity rel $E$ if and only
if the physical braid $\bb$ describing the motion
of $E$ is \wonko\ to $\trivb$. Here is the proof.
If $\bb$ is \wonko\ to $\trivb$ by the family
$\psi_t$, then $h_t'= \psi_t\circ h_t$ is a fluid motion
that fixes $E$ for all $t\in[0,1]$ and $h_1' = h_1=f$,
and so $f$ is isotopic to the identity. Conversely, 
if $h_t'$ is a fluid motion that 
fixes $E$ for all $t\in[0,1]$ and $h_1' = h_1=f$, then
$\psi_t = h_t'\circ h_t\inv$ is a family that will
make $\bb$ \wonko\ to $\trivb$.

The next step is to see how the braid word describing  the motion of $E$ 
indicates when the advection \diffeo\ is isotopic to the
identity rel $E$. This is certainly  the case  when the mathematical 
braid is the identity element in the braid group, but this is
not the only case. Consider the plane
motion given in complex coordinates by $h_t(z) = |z| \exp(2 \pi i t)$.
 For a given $n$, let $E$ be the set of $n^{th}$ roots of unity, 
$\{ \exp 2\pi i j/n:j = 1, \mydots, n\}$.  Then
it is easy to check that the physical braid $\bb$  of the motion of
$E$ has a mathematical braid 
$(\bs_1\bs_2\mydots\bs_{n-1})^n$ and further that the \diffeo\ 
$\Psi(z,t) =   (|z| \exp(-2 \pi i t), t)$ makes $\bb$ \wonko\ to $\trivb$.
It is known  that this  mathematical braid and its
powers are the only cases where this happens ([Bm1]). More
precisely, if we let $\hB_n$  be $B_n$  with the added  relation that
$(\sigma_1 \sigma_2 \mydots \sigma_{n-1})^n$ is the identity
element, then the braid word of $\bb$ is equivalent to the identity
element of $\hB_n$ exactly when $\bb$ is \wonko\ to $\trivb$. 
Note that the center of $B_n$ (\ie all the elements which
commute with every other element) is exactly all the powers
of $(\sigma_1 \sigma_2 \mydots \sigma_{n-1})^n$, so  one could more
succinctly define $\hB_n$ as $B_n$ mod its center.

For $n>2$,  $\hB_n$  has infinitely many
elements, and so there are many \diffeos\ that are not isotopic to the
identity; the simplest example is when $E$ is three points and
$F$ is a rel $E$ \diffeo\ with mathematical braid $\sigma_1^2$. 
When \diffeos\ are not isotopic to the identity, isotopy
can be described using the identity case as follows.
  Two rel $E$  \diffeos\ $f$ and $g$  are
 isotopic if there is third \diffeo\ $h$, with  
$h$ isotopic to the identity, and $g = h \circ 
f$. If  $f$ and $g$  are time one  maps of fluid motions, this
says that they are  isotopic if  we can accomplish the same 
advection as $g$ by first allowing the advection for
$f$ and then following it by a fluid motion
that keeps the points of $E$ fixed. 

The  connection of  braids to this more general notion of isotopy 
requires the general definition of  \wonko\ physical braids.
 Two physical braids $\bb^0$ and $\bb^1$, both with  endpoints $E$ 
 are \wonko\ if there is a level preserving
 \diffeo\ $\Psi:\reals^2\times[0,1]\raw\reals^2\times[0,1]$ 
 that takes  $\bb^0$ to $\bb^1$ where 
 $\Psi(z, t) = (\psi_t(z), t)$ for some family
of \diffeos\ $\psi_t$ with $\psi_0=\psi_1 = id$. 
 Using the result in the identity case, we have that  two rel
$E$ advection \diffeos\ are isotopic if and only
if their  physical braids with endpoints $E$ are \wonko\ which
happens if and only if the corresponding braid words are equivalent
in $\hB_n$. In different language, the collection of rel
$E$ isotopy classes forms a group under composition; this
group is isomorphic to $\hB_n$, where $n$ is the number of 
elements in $E$.

It is worth noting that when two physical braids with 
endpoints $E$  are
equivalent in the sense defined in \S 4.1, they
are \wonko\ (hence the use
of the modifier ``weakly''). This is easily seen using the
corresponding mathematical braids. The equivalent physical
braids have the same braid word in $B_n$, and hence in 
$\hB_n$, and thus they are weakly equivalent. The converse
need not be true, the simplest example being the trivial braid
and a physical braid having braid word 
$(\sigma_1 \sigma_2 \mydots \sigma_{n-1})^n$. These represent the
same word in $\hB_n$, but not in $B_n$, and hence
the physical braids are \wonko\ but not equivalent.

The last step is to  extend  these ideas to allow different distinguished sets
$E$ so that vortex motions with different initial configurations
can be compared.
A rel $E$ \diffeo\ $f$ and a rel $E'$ \diffeo\
$f'$ are {\de  isotopic up to conjugacy } if there is
a \homeo\ $h$ with $h(E') = E$ and $h f' h\inv$ is isotopic
to $f$ rel $E$. Thus $f$ and $g$ are isotopic after
a change of coordinates. 
Two physical braids $\bb$ and $\bb'$ with endpoints  
$E$ and $E'$  are \wonko\ if there exists a \diffeo\
$h:\reals^2\raw \reals^2$ with $h(a_i(0)) = a_i'(0)$ for all
$i$ (and so $h(E) = E'$) and the physical braid $\{ (h(a_i(t)), t)\}$
is \wonko\ to $\bb'$. We then have that 
 two advection \diffeos\ are  isotopic up to conjugacy
if and only if their physical braids are \wonko, which happens 
if and only if their mathematical braid words are conjugate in $\hB_n$.

\medskip

\noindent{\bf \S 5.2  The Thurston-Nielsen representative.}
The Thurston-Nielsen theory 
provides tools to analyse \homeos\ using their isotopy classes. 
The theory constructs  in each isotopy class  a 
special \homeo\ which is now called the Thurston-Nielsen 
(TN) representative.
This TN-\homeo\ is in a precise sense the simplest map in the
isotopy class. Simplest here means in both the topological and 
dynamical sense. Thus once we understand the topology and dynamics 
of the TN-\homeo\ in an isotopy class, we know 
dynamical and topological complexity that must be present
in {\it every} \homeo\ in the class. 
This subsection and the next contain an  introduction 
to the \TN\  theory that  is targeted for our  applications.
In particular, only \homeos\ of the punctured plane are considered here,
or equivalently, \homeos\ of the plane rel a finite set $E$.
The \TN\  theory holds for the much wider class of 
surfaces of negative Euler characteristic.
The reader is urged to consult  [T], [FLP] or [CB] for a more balanced and
complete  treatment. For a survey of the dynamical applications
of the \TN\ theory, see [Bd].

The TN-\homeos\ are of three basic types. The first type  consists
of  dynamically very simple maps  called {\de finite order (fo)}. These 
are defined by 
the property that for some $m>0$ the map composed with itself $m$ times 
equals the identity map. 
The second type of TN-\homeos\ are dynamically very complicated and
are called {\de pseudoAnosov (pA)} \homeos. These will be discussed 
in greater detail in the next subsection because the presence of a 
pA \homeo\ in an isotopy class has strong implications for the dynamics of
an advection \homeo.
The final type  of TN-\homeos\ are {\de reducible}, which means that 
there is a collection of simple disjoint loops 
with the property that the loops are permuted by the
reducible \homeo. Cutting along the loops yields
a collection of smaller surfaces and on each of these 
surfaces the reducible \homeo\ is either finite order or
pseudoAnosov.

A given isotopy class can only contain a TN-map of one type,
and thus an isotopy class is called  pseudoAnosov,
finite order or reducible depending on what kind of TN-map it contains.
Note that  the conjugate of a TN-map is also a TN-map of the
same type.
Thus, using the results of the previous subsection, one may
also speak of the TN-type of a braid or a braid type.
There are many theorems which help decide the TN-type of
a braid or isotopy class.  Rather than catalog these results here
it will suffice to note that there  is a computer implementation of 
an algorithm due to Bestvina and Handel ([BH]) (cf. [FM] and [Ls])
which, given the braid word,
decides the TN-type. In the pA case the algorithm outputs the
isotopy invariant dynamical data described in the next
subsection.\footnote{$^1$}{There is a C++ implementation of this algorithm
 by Toby Hall (with a Win95 interface) available for download. Contact
the first author for site locations.}

 For planar regions the finite order braids and rel $E$ isotopy
classes are well known. If we let $R_{n}$
denote rigid rotation of the plane by $-2 \pi /n$, then certainly
$R_n^n = id$, and so $R_n$ is finite order. 
We may consider $R_n$ as the time one flow map of the
motion  given in complex coordinates by $h_t(z) =- |z| \exp(2 \pi i t/n)$.
If $E$ is the set of $n^{th}$ roots of unity, 
its physical braid under this motion is 
 $\beta_n := \sigma_1 \sigma_2 \mydots \sigma_{n-1}$. 
Note that $\beta_n^n$ is the identity element of $\hB_n$, and 
making this property hold is, in fact, the defining feature of $\hB_n$.
Thus $\beta_n$ is a \fo\ braid and so is   $\beta_n^m$ for any 
$0\leq m \leq n-1$.  A classic result of Brouwer (see section 8.2 in
[Bd]) says that all
\fo\ \homeos\ of the plane are conjugate to  rigid rotations.
This implies that the $\beta_n^m$ are the only finite order
braids on $n$ strands.  Examples of other braids with various TN-types
are given with vortex examples in \S 6.2. 

For planar regions the \TN\ theory only has content
for regions having three or more punctures or holes, or equivalently,
for isotopy classes rel three or more points. When $E$ contains
just two points, all  TN-\homeos\ are finite order
since $\hB_2$ only contains the identity element $e$ and the braid
$\sigma_1$ with $\sigma_1^2 = e$.  Classic
results of Alexander (see Cahpetr XV in [D]) say that when $E$ has one or
zero elements, there is only one isotopy class; all
orientation preserving \homeos\ of the plane are
isotopic, as are those of the once punctured plane.

\medskip
\noindent{\bf \S 5.3  PseudoAnosov maps and isotopy classes.}
Linear Anosov maps on tori provide one of the best known and understood
examples of chaotic dynamics. The
\homeo\ $\phi$  of  the two torus induced by the matrix
$$M = \pmatrix{2 & 1\cr 1 & 1\cr}$$
acting on the plane is  called 
Thom's toral automorphism or Arnolds' cat
map,  the latter nomenclature derived from  an illustration in [AA].
The \homeo\ $\phi$ has a Markov partition with 
transition matrix $M$  that
gives a symbolic description of the dynamics.
The largest eigenvalue $\lambda>1$ of $M$  controls much
of the dynamical asymptotics. 
The number of fixed points of $\phi^n$ grows like 
$\lambda^n$ as does the length of topologically nontrivial loops
under iteration.   In addition, 
these properties are shared in the appropriate sense by any \homeo\
on the torus that is isotopic to $\phi$ ([Fk]).

For the map $\phi$ there is uniform stretching and contraction by $\lambda$ at 
every point, and the corresponding eigendirections  fit together to yield a 
pair  of foliations of the torus by unstable and stable manifolds. 
Such nonsingular foliations cannot live on surfaces of negative
Euler characteristic like the multi-punctured plane, and thus these
surfaces cannot support Anosov maps.  However, pseudoAnosov
maps do exist on these surfaces and share many of the properties
of Anosov maps.

In  pseudoAnosov maps there is still uniform stretching 
and contracting at each point by a  factor $\lambda$.
The unstable and stable directions still fit together into
invariant  unstable and stable foliations  but
these foliations must  have  
a finite number of points (called singularities or prongs) where 
there is either one  or three or more stable directions and 
the same number of unstable directions.
Despite the presence of these singularities,
pseudoAnosov maps  share most of  the  basic properties
of Anosov maps.  There is a Markov partition with transition matrix $M$ that
allows one to code the dynamics of the pA map $\phi$.  The largest
eigenvalue of $M$  is the stretching factor $\lambda>1$.
The number of fixed points of $\phi^n$ grows like 
$\lambda^n$ as does the length of topologically nontrivial loops
under iteration. The precise way in which these properties are
shared by any isotopic map is described by the following theorem
due to Handel ([H]). The notation $g_{|Y}$ means the map $g$ restricted
to the set $Y$.

\medskip
{\bf Theorem:} {\it If $\phi:M^2\raw M^2$
is a pseudoAnosov \homeo\ and $g$ is isotopic to $\phi$,
then there exists a closed, $g$ invariant set $Y$, and a continuous,
onto map $\alpha:Y\raw M^2$, so that $\alpha\circ g_{|Y}= \phi\circ\alpha$.
Further, for any \pp\ $x$ of $\phi$, there is a \pp\ of the same 
period $y$ of $g$ with $\alpha(y) = x$.} 

\medskip

Thus any map $g$ isotopic
to a pA map has a compact invariant set that is semiconjugate
to the pA map.  Since the map $\alpha$ is onto, the dynamics
of $g$ are at least as complicated as those of the pA map $\phi$.
Thus, for example, the exponential growth rate of the number of fixed points
of $g^n$ is at least $\lambda$
and the topological entropy
of $g$ is  greater than or equal to $\log(\lambda)$.
Note that nothing prevents the
dynamics of $g$ from being more complicated than that of $\phi$, the theory
provides just a lower bound.

A loop in a surface is called topologically nontrivial
if it  cannot be deformed into  a point,  a puncture, or
a boundary component.  Another basic result about pA maps
is that the length of topologically nontrivial loops grows like
$\lambda^n$ under iteration. Via the semiconjugacy, this implies
a similar growth under  $g$. It is important to note that this
does not imply that the map $g$ has Lyaponov exponents equal
to $\log(\lambda)$ everywhere.  One does obtain from
smooth ergodic theory  that $g$ has an
ergodic invariant measure with an exponent at least $\log(\lambda)$
(cf. [KH]), but this measure will be supported on the set $Y$,
which could be small with respect to the usual measure on $M$.

Since the number $\lambda$ can be  computed by the \BH\ algorithm,
the braid describing an isotopy class yields
quantitative information about any \homeo\ in the
isotopy class.  An example
of an isotopy class containing an
advection \homeo\ induced by vortex motion is given \S6.3.
The \BH\ algorithm also returns a one dimensional graph
called a {\de train track} in addition to a self-transformation of
the graph. The edges of the graph form the   Markov
partition for the pA map, and the self-transformation of
the graph provides the transition matrix as well
as the structure of the invariant foliations. These foliations
are sometimes called the ``invariant manifold template'' of
the pA map. The semiconjugacy then gives a lower bound or skeleton for  the
invariant manifold templates of $g$.

\medskip

\noindent{\bf \S 6 Advection and isotopy classes.}

In this section we use the \TN\ theory to
study the of advection of a passive particle
in the velocity field produced by three point vortices.
In \S 6.1 a Poincar\'e map (or advection \homeo) is defined whose
 iterates describes the dynamics of 
 passive scalars. This is a standard construction, but
an adjustment is required
because of  the  dynamic phase in the periodic motion of the vortices.
  In \S6.2 we study isotopy classes of the advection \homeos\ in the plane 
and array case using the braids computed in \S 4.
 Since all the motions in a single regime
have the same braid,  all the corresponding advection \homeos\ 
have the same TN type. For the planar case, these are all
 reducible or \fo, but the array case has regimes of pA type.
In \S6.3 we study one such pA regime in detail.

\medskip
\noindent{\bf \S6.1 The Poincar\'e map.} 
As we have seen, a vortex motion $z_\alpha(t)$ can be obtained by solving
 (2.1) directly, or as in \S 3, by  using (3.4) 
and a period $P$ solution $Z(t)$ of the reduced equations (3.5).
The vortex motion, in turn, generates a velocity
field given by the right hand side of (2.3).  The goal of  this subsection
is to define a Poincar\'e map. The  iterates of this map will describe
the time evolution of passive particles advected in this
velocity field.  There is a standard construction of
such a map for an equation  of 
 the form $\dot z = F(z, t)$ with $F$ a velocity field $P$-periodic 
in $t$.  The time-$P$ flow map $f$ as defined
in \S5.1 by $f(z_0) = z(P; z_0, 0)$ will  satisfy
$f^n(z_0) =  z(nP; z_0, 0)$, and so the iterates of $f$ describe the
time evolution of points in discrete steps of size $P$.

For the advection caused by the point vortices
this construction needs to be altered slightly
because the vortex trajectories satisfy $z_\alpha(t + P) = z_\alpha(t) + b$, 
with $b$ the dynamic phase associated with $Z(t)$. The easiest
alteration is to pass to a uniformly translating frame with
velocity $b/P$.   If $F(z,t)$ denotes the right hand side of (2.3) 
corresponding  to the given vortex motion, define
the {\de advection \homeo} $f$ as the 
time $P$-flow map of the equation 
$$\dot\zeta = F(\zeta + {b\over P} t, t ) - {b\over P}.$$
This advection \homeo\ satisfies $f^n(z_0) = z(nP;z_0, 0) - nb$,
and so  $f$ describes the position of a particle
after advecting for one
period $P$ and then translating by $-b$. Alternatively, this
could have been taken as the definition of $f$, \ie 
$f(z_0) = z(P;z_0, 0)-b$ with $z$ the solution of (2.3).
Points that are periodic under $f$ correspond
to advected particles that are translating along with
the vortices, but after some integer multiple of $P$ they
have returned to their initial positions relative to
the vortices.
Although the velocity field is
not defined at the positions of the vortices, the \homeo\
$f$ can be extended (continuously, not differentially) to
make the three initial positions of the
vortices, $\za(0)$,  fixed points.  

In \S 4.3 the motion of a vortex array in  the
cylinder was transformed to the plane minus the origin by the conformal map
$T$. The \adhom\ in this case is transformed to
${\hat f} = T f T\inv$. Note that ${\hat f}$ may be extended to make 
$0$ a fixed point.  Thus 
all the $s_\alpha(0)$ from (4.4) can be considered as fixed points 
of ${\hat f}$.

\medskip
\noindent{\bf \S6.2 The TN type of regimes.}
As seen in \S 4.2 and \S 4.4, all the \pos\ in the same regime in the reduced
plane give rise to  the same braid. By results in \S4.1 this implies that all
the corresponding advection \homeos\ have the same braid type, which  
allows us to speak of the TN type of a regime. If this
type is pA, then as in \S5.3 all the advection \homeos\ will share 
a certain set of dynamics. 

In the case of three vortices in the plane, only two
braid types arose as descriptions of regimes.
In one of them, a pair of vortices
rotate around one other while the other vortex is uninvolved.
This is a reducible class, with a finite order component containing the
pair of interacting vortices. The other regime, D, had braid
$(\sigma_1 \sigma_2)^3$, which is the identity element in
$\hB_3$, and the advection \homeo\ in this case is isotopic
to the identity, with the isotopy consisting of
unwrapping the plane via single rotation; this topologically
undoes whatever advection was caused by the vortices. 
Since only pA classes or components yield  dynamical
information from the \TN\ theory, the theory gives no
information about advection dynamics in the planar case.

The \TN\ theory does provide dynamical information about
advection induced by
vortex arrays. The Bestvina-Handel algorithm was
used to compute the TN type of the various 
regimes from Figure 3.2. Using the braids given in the second
column as input, the computed TN type is shown in the 
third column of Table 4.1.   When a regime is pA (or has a pA component),
the expansion constant is given in the right column. 

As we comment on the various regimes the reader
is urged to draw a picture of the appropriate braids. 
The regimes which contain poles, III through VIII, correspond to a pair
of vortices which rotate  around one other while the
other is uninvolved. As in the planar case, 
the isotopy classes of the advection \homeos\
are reducible  with a finite order component containing the
pair of interacting vortices. 

In the braid for regime X shown in Figure 6.1a, 
 the  three right most strands 
do not link with the left most string. This shows that  the 
corresponding isotopy class is
reducible with a reducing circle containing the three right most 
strands.  Inside this reducing circle the behaviour as a 
{\it three}  braid is
 $\sigma_2 \sigma_1\sigma_2 \sigma_1\sigma_2\sigma_1 =
(\sigma_1\sigma_2)^3$ using the relations in $B_3$. This braid  is the
same as the braid for region D in the planar case, which represents
the identity class. Thus for regime X the isotopy classes
are reducible with all finite order components. 
The braid for regime IX is reducible in a  way similar to
that of regime X, 
but inside the reducing circle the class is again reducible, but
still with all \fo\ components. 

\midinsert
\def\epsfsize#1#2{.8\hsize}
\centerline{\epsfbox{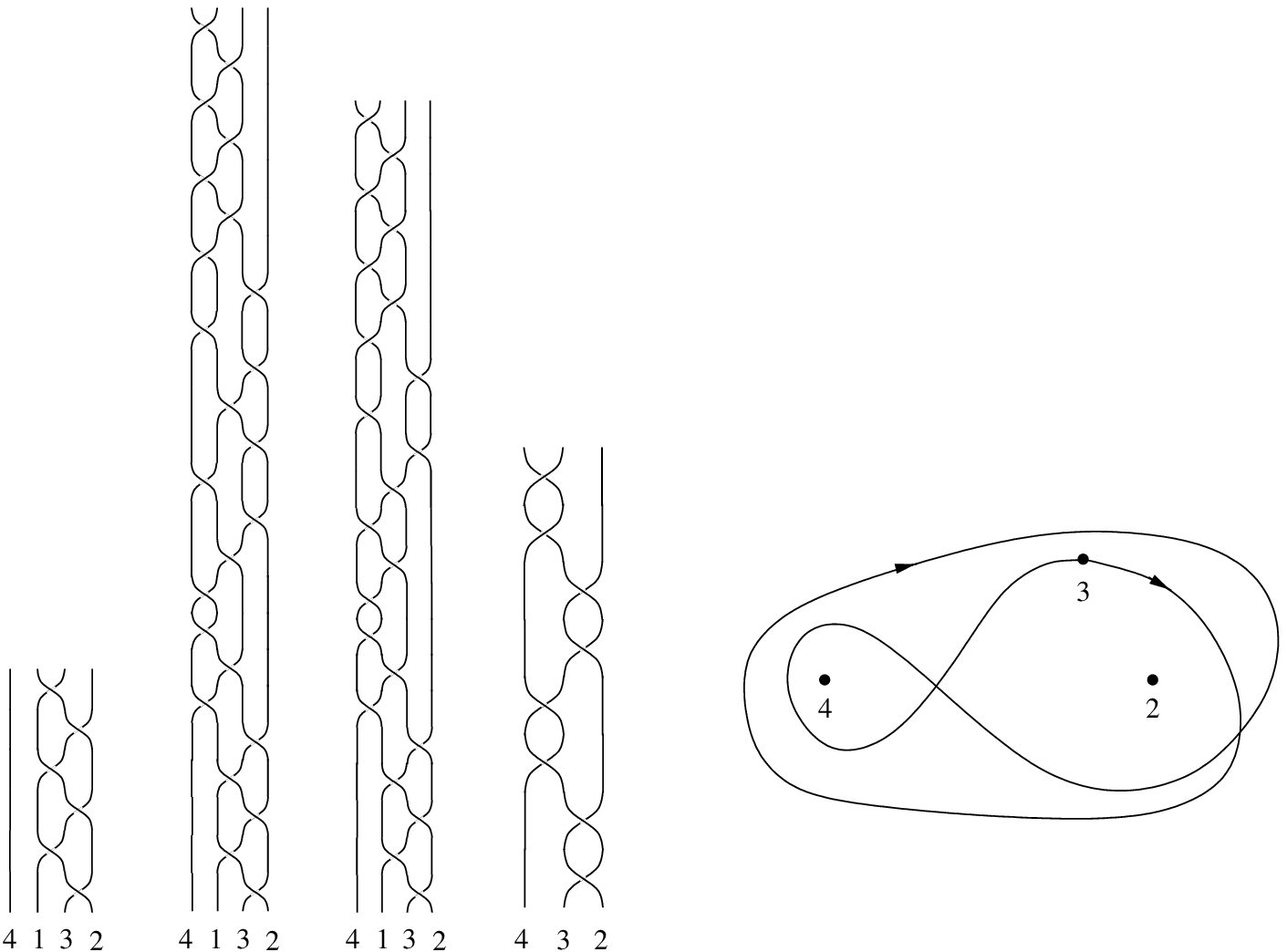}}
\medskip
{\leftskip=60pt\rightskip=60pt\parindent=0pt

{\bf Figure 6.1:}
Mathematical braids of regimes from Figure 3.2: (a) Regime X
(b) Regime XII (c) Regime XI. The braid of regime XI restricted
to vortices 2, 3, and 4 is shown in (d), and a  schematic
of the plane motion of the vortices is shown in (e)

}
\endinsert

In the braid for regime II shown
in Figure 4.4d. The schematic of the plane motion in Figure 4.4b makes
it clear that the class is reducible; one can enclose the motions
in successive, nested  reducing curves.
The classes for regime I are similar.

The braid for regime XII is shown in Figure 6.1b. 
The braid makes it clear that 
 vortices 1 and 3 have essentially the same motion with
respect to vortex 2 and the origin (``vortex'' 4). 
They do rotate around each 
other once, but they may be enclosed in a reducing circle. 
 To analyse the class outside the reducing circle we
 think of squeezing the strands of
 vortices $1$ and $3$ into a single strand,
 or equivalently, we delete one of the strands.  The
 result is the  three braid
 $\s_2^2\s_1^2\s_2^2\bs_1^2$  shown in Figure 6.1d. 
This braid corresponds to a pA class  rel $3$ points
 with $\lambda \approx 13.9$. 

Regime XI also yields a pA class with  $\lambda \approx 13.9$. 
Although this class is irreducible, the strand corresponding to
vortex 1 is in fact a fixed point of the pA map in the class defined by
the other three strands. The latter is the same as that for
regime XII (see Figure 6.1).  The dynamics in regime XI are considered in
detail in the next subsection. 
The last regime is XIII which is pA with $\lambda\approx 9.9$.

\medskip
\noindent{\bf \S6.3 An example of the dynamical implications of the \TN\ theory.}
As an example we describe
some of the results afforded by the \TN\ theory for
advection \homeos\ coming from vortex motions in  regime XI.
Recall that this regime's TN-type is pA
with $\lambda \approx 13.9$. By fixing a
 \po\ $Z(t)$ in this regime and an initial
position $z_1(0)$ we obtain a motion of all
the vortices as in \S6.1. The corresponding advection
\homeo\ is denoted $f$.
 We may variously consider $f$ as a \homeo\ of the singly-periodic
plane, the cylinder, or as transformed from the cylinder
to the plane by the conformal map $T$.

Using the theorem of \S5.3 and the comments after it we know that
the number of fixed points of $f^n$ grows with $n$ at least like
$13.9^n$. These fixed points for $f^n$ correspond in the lab frame
to passive particles that are advected by the vortex velocity
field and after $n$ periods have returned to the same
position {\it relative to the vortices}. They have thus translated by
$n b$, where $b$ is the relevant dynamic phase.
We also know that under advection, topologically nontrivial
curves grow in length at least at a rate of $13.9^n$. In addition, the
topological entropy of $f$ is at least $\log(13.9)$. This implies
that $f$ has  a hyperbolic invariant measure (in the sense of 
[KH]) with Lyapunov exponents
$\log(13.9)$, but no information is available regarding its support.

As noted at the end of \S4.2,  although all the
advection \homeos\ arising from vortex motions in regime
XI will have these properties,
the time scale of the advection described by the
\homeo\ will vary greatly amongst  motions in the same regime.
In addition,
variations in the dynamic phase will make a significant difference
in the motion as observed in the lab frame.

The picture of the trajectories of the vortices  transformed into the
plane  is also instructive. Figure 6.1c shows the mathematical 
braid of this regime. Figure 6.1d shows just the motion of
vortices  3 and 2 and the origin ($=$ ``vortex'' 4),
while Figure 6.1e shows a schematic
of this motion in the frame of the vortex 2 after it has
been transformed to the plane. The feature of note is that
the trajectory of vortex 3 crosses itself in an
essential way. We cannot remove this self-intersection
by  deforming the path in the complement
of vortex 2 and the origin. This essential self intersection
in fact implies that the isotopy class of the advection \homeo\
generated by vortex 3, vortex 2  and the origin is pA.
The self-intersection forces  regions of the fluid 
to be repeatedly pushed across pieces of themselves
as they are advected. This forces the  advection \homeo\
 to constantly stretch and fold in an essential
topological fashion. It is well known that stretching and
folding is a basic mechanism for the generation of chaos
and in this case it is topological in nature and  is present in
any map in the isotopy class.
\medskip

\noindent{\bf \S7 Summary and conclusions.}

Three-vortex systems with zero net circulation have shown themselves
to be worthy of both  physical and mathematical interest.
Using  a combination of dynamical, geometric and topological
methods  a fairly complete analysis of these systems has been achieved.
After fixing
a value of the integrals, the reduction process maps the system onto a one
degree of freedom Hamiltonian. This Hamiltonian system has a natural subdivision
into regimes and we have seen that all the solutions corresponding to
a single regime yield  topologically identical vortex motions as 
described by their braids.  Viewing the vortices as the  ``large scale motion'' 
the use of braids allows one
to make precise sense of the topology and structure of 
the large scales.
In addition, the notion of reducibility of a braid or
isotopy class  allows  precise description
of the  hierarchy within the large scale motions.
For certain regimes  the \TN\ theory uses the topology of
the vortex motion to obtain 
very strong conclusions about the advection caused by
the large scales. Although point vortices describe a very idealised
fluid, it certainly of value  to have model systems for which  
commonly used phrases such as ``structure'' and  ``large scale''
have a precise  meaning with computable implications.

Although this paper has focused on a particular kind of vortex
system it should be
clear that the  methods employed have much wider applicability. 
Any system involving the periodic  motions of points in a
surface has a natural description using braids. 
If there is a natural \homeo\  on  the complement of the points,
then it can be studied using \TN\ theory. The prototypical such
system is an iterated \homeo\ of a surface in which case the
collection of points is a collection of \po s. This application is
fairly well studied, see [Bd] for a survey. A closely related
example is a periodically forced oscillator, see  [McT] for
a survey.  Within fluid mechanics, a two dimensional
fluid region with a periodic stirring protocol by rods
is studied from this point of view in [BAS].

It is also clear that the methods of this paper apply to 
a wider class of vortex systems. 
For example, the methods presented
here immediately apply to the periodic motions of $n>3$
vortices. However, in the $n>3$ case, the reduced dynamics is 
described by a $n-2 > 1$ degree of freedom Hamiltonian system.
In this case one expects that the generic bounded orbit is not periodic, but
rather only recurrent (via the Poincar\`e Recurrence Theorem).
At a time of near periodicity, such an orbit can be closed
in a canonical way and its braid computed. The corresponding
advection \homeo\ may then be studied using the  \TN\ theory. 
The theoretical task is then to understand how the conclusions
derived from  the closed motion can be  applied to the advection
caused by actual recurrent motions.  Similar methods could be then
applied to the motions of regions of very concentrated
vorticity in two-dimensional turbulence and the resulting fluid mixing.

A distinct advantage of the  application of the \TN\ theory 
is that it gives quantitative, {\it a priori} lower bounds for the chaotic
dynamics based only on qualitative, topological data. 
However, the methods are only applicable after a punctured plane
is obtained, and then the \homeo\ under question must
be in a \pA\ isotopy class. When the methods are applicable
an important question is the relative importance of the 
dynamics dictated by the \TN\ theory  compared to  other sources
of chaos. For example, in advection due to point vortices, the
singular vorticity gives rise to very violent 
local stretching.   As the vortices move this highly stretched fluid
can fold over itself in the classic ``chaos engine'' scenario. 
While this effect can be deformed away by an isotopy 
relative to the vortices, preliminary
numerical investigations indicate that in certain cases
this effect is an order of
magnitude greater than the lower bounds given by \TN\  theory. For 
a smooth distribution of vorticity, this local stretching
effect would diminish,
and so the global effects given by \TN\ theory would presumably
gain more importance. 

An example where the \TN\  effects are clearly
dominant is the  slow stirring 
of a viscous fluid studied in [BAS]. Three stirring
rods in glycerine were moved in a protocol corresponding to  pseudoAnosov braid.
Dye tracers indicate  an advection dynamics very close to
the pA model. The  structure of 
the invariant manifold template of the pA map is  clearly
visible. The fluid motion in this case seems to realize
the least dynamical complexity allowed in its isotopy
class which corresponds to   very efficient stirring.

The topological methods employed here are very successful in
illuminating certain features of the vortex system. 
As with any method, there are situations of 
greater and lesser applicability, but when applied with 
care these topological techniques  provide a powerful tool in the 
study of two-dimensional chaotic dynamics.

\bigskip
\centerline{\bf References}
\medskip

\item{[AR]} Adams, M.  and Ratiu, T., The three-point vortex 
problem: commutative and noncommutative integrability,   
{\it  Contemporary Mathematics}, {\bf 81}, 1988, 245--257.

\item{[Af1]} Aref, H.,  Integrable, chaotic and turbulent vortex motion in
two-dimensional flows, {\it Ann. Rev. Fluid Mech.}, {\bf 15}, 1983, 345-389.

\item{[Af2]} Aref, H., Three-vortex motion with zero total circulation:
 Addendum (Addendum to paper by N. Rott),
{\it  J.  Appl. Math. Phys. (ZAMP)}, {\bf  40}, 1989,
 495--500. 

\item{[AS1]} Aref, H. and Stremler, M. A., 
On the motion of three point vortices in a periodic strip,  {\it 
J. Fluid Mech.}, {\bf  314},  1996, 1--25. 

\item{[AS2]} Aref, H. and Stremler, M. A.,
 Vortex quasi-crystals, in preparation.

\item{[A1]} Arnol'd, V. I.,  Remarks on quasicrystallic symmetries,
  {\it Phys.  D}, {\bf  33},  1988, 21--25. 

\item{[A2]} 
 Arnol'd, V. I., {\it  Mathematical methods of classical mechanics},
Springer-Verlag, 1989.

\item{[AA]} Arnold'd, V. I. and Avez, A., 
{\it Ergodic problems of classical mechanics}, Benjamin, 1968.

\item{[BH]} Bestvina, M. and Handel, M.,  Train tracks for surface
 homeomorphisms, {\it Topology}, {\bf 34}, 1995, 109--140.

\item{[Bm1]} Birman, J, {\it Braids, Links and Mapping Class Groups},
 Annals of Mathematics Studies, Princeton University Press, 1975.

\item{[Bm2]} Birman, J., On braid groups, {\it Com. Pure and
Appl. Math.}, {\bf 22}, 1969, 41--72.

\item{[BL]} Birman, J. and Libgober, A., ed., 
Proceedings of the AMS-IMS-SIAM Joint Summer Research Conference 
on Artin's Braid Group, {\it  Contemp.
Math.}, {\bf 78}, 1988.

\item{[Bd]} Boyland, P. Topological methods in surface dynamics,
 {\it Topology and its Applications}, {\bf 58}, 223--298, 1994.

\item{[BAS]} Boyland, P., Aref, H. and Stremler, M., Topological
fluid mechanics of stirring, {\it J. Fluid Mech.}, in press.

\item{[CB]} Casson, A. and Bleiler, S., {\it Automorphisms of Surfaces
after Nielsen and Thurston}, London Math. Soc. Stud. Texts, {\bf 9},
Cambridge University Press, 1988.

\item{[C1]} Chorin, A. J., Turbulence and vortex stretching on a lattice,
 {\it Commun.  Pure Appl. Math.}, {\bf 39} (Supplement), 1986, S47--S65.

\item{[C2]} Chorin, A. J., Scaling laws in the vortex lattice model 
of turbulence, {\it   Commun. Math. Phys.}, {\bf 114}, 1988, 167--176.

\item{[C3]} Chorin, A. J., Spectrum, dimension, and polymer analogies in fluid
turbulence, {\it  Phys. Rev. Lett.}, {\bf  60}, 1988, 1947--1949.

\item{[D]} Dugundji, J., {\it Topology}, Allyn and Bacon, Inc., 1966.

\item{[FLP]} Fathi, A., Lauderbach, F. and Poenaru, V., Travaux
de Thurston sur les surfaces, {\it  Asterique}, {\bf 66-67},  1979.

\item{[Fk]} Franks, J., Anosov diffeomorphisms,
{\it AMS Proceedings of Symposia in Pure Mathematics},
{\bf XIV}, 1970,  {61--93}.

\item{[FM2]} Franks, J. and Misiurewicz, M., Cycles for disk
homeomorphisms and thick trees, 
 {\it Contemp. Math.},  {\bf 152}, 1993, 69--139.

\item{[H]} Handel, M., Global shadowing of pseudo-Anosov homeomorphisms,
 {\it Ergod. Th. \& Dynam. Sys.}, {\bf 5}, 1985, 373--377.

\item{[J]} Janot, C.,  {\it Quasicrystals: a primer},
 Oxford University Press, 1992. 

\item{[KH]} Katok, A. and Hasselblat, B., {\it Introduction to
the modern theory of dynamical systems}, Cambridge University Press, 1995.

\item{[L]} Leonard, A., Computing three-dimensional incompressible flows with
vortex elements, {\it  Ann. Rev. Fluid Mech.}, {\bf  17}, 1985, 523--559.

\item{[Ls]} Los, J., Pseudo-Anosov maps and invariant train
tracks in the disc: a finite algorithm, {\it Proc. London Math. Soc.},
{\bf 66}, 400--430, 1993.

\item{[McT]} McRobie, F. A. and Thompson, J. M. T., Braids and knots
in driven oscillators, {\it Intern. J. Bifurcation \& Chaos},
{\bf 3}, 1993, 1343--1361.

\item{[Mj]} Majda, A. J., Vorticity, turbulence, and acoustics in fluid flow,
 {\it SIAM Rev.}, {\bf 33}, 1991, 349--388.

\item{[M]} J. Marsden {\it Lectures on Mechanics}, LMS Lecture Note
Series, {\bf 174}, Cambridge University Press, 1992.

\item{[MZ1]} Marsden, J. E., O'Reilly, O. M., Wicklin, F. J., and
 Zombro, B. W.,  Symmetry, stability, geometric
phases, and mechanical integrators I, {\it  Nonlinear Sci. Today},
{\bf  1},  1991, 4--11. 

\item{[MZ2]} Marsden, J. E., O'Reilly, O. M., Wicklin, F. J., and
 Zombro, B. W.,  Symmetry, stability, geometric
phases, and mechanical integrators II, {\it  Nonlinear Sci. Today},
{\bf  2},  1991, 14--21. 

\item{[MH]} {Meyer, K.R. and  Hall, G.R.},
 {\it Introduction to Hamiltonian Dynamical Systems and the N-Body
 Problem}, Springer-Verlag, {1992}.

\item{[MfT]} Moffatt, H. K. and  Tsinober, A.,
 Helicity in laminar and turbulent flow,
{\it Ann. Rev. Fluid Mech.}, {\bf 24}, 1992, 281--312.

\item{[Mt]} Montgomery, R., 
 The $N$-body problem, the braid group, and 
action-minimizing periodic solutions,
{\it Nonlinearity}, {\bf 11},  1998, 363--376.

\item{[O]}
O'Neil, K. A.,  On the Hamiltonian dynamics of vortex lattices,
{\it  J. Math. Phys.}, {\bf 30}, 1989, 1373--1379. 

\item{[PS]} Pullin, D. I.  and  Saffman, P. G.,
 Vortex dynamics in turbulence, {\it   Ann.  Rev. Fluid Mech.}, {\bf  30},
 1998, 31--51.

\item{[R]} Roberts, P. H., A Hamiltonian theory for weakly interacting vortices,
{\it Mathematika}, {\bf 19}, 1972,  169--179.

\item{[S]} Saffman, P. G., Dynamics of vorticity, {\it  J. Fluid Mech.},
{\bf 106}, 1981, 49--58.

\item{[SB]} Saffman, P. G. and  Baker, G. R.,  Vortex interactions, 
{\it Ann. Rev. Fluid Mech.}, {\bf 11}, 1979, 95--122.

\item{[Srp]} Sarpkaya, T., 
Computational methods with vortices - The 1988 Freeman
Scholar lecture, {\it  J. Fluids Engin.}, {\bf 111}, 1989, 5--52.

\item{[SL]} Shariff, K. and  Leonard, A.,  Vortex rings,
{\it  Ann. Rev. Fluid Mech.}, {\bf 24}, 1992, 235--279.

\item{[SA]} Stremler, M. A. and Aref, H., 
Motion of three point vortices in a periodic
parallelogram, {\it J. Fluid Mech.},  to appear.

\item{[T]} Thurston, W., On the geometry and dynamics of diffeomorphisms of
surfaces, {\it  Bull. A.M.S.}, {\bf  19}, 417--431,  1988.

\item{[Z]} Zabusky, N. J., Computational synergetics and mathematical
 innovation, {\it  J. Comp. Phys.}, {\bf 43}, 1981, 195--249.

\bye